\documentclass[preprint,a4paper,showabstract,superscriptaddress ]{revtex4-2}
\setcitestyle{super}
\usepackage{multirow}
\usepackage{bm}

\usepackage{graphicx}
\usepackage{epstopdf}
\usepackage{dcolumn}
\usepackage{bm}
\usepackage{hyperref}
\usepackage[utf8]{inputenc} 
\usepackage[T1]{fontenc}    
\usepackage{lmodern}
\usepackage{url}            
\usepackage{booktabs}       
\usepackage{amsfonts}       
\usepackage{nicefrac}       
\usepackage{microtype}      
\usepackage[version=3]{mhchem}
\usepackage{natbib}
\usepackage{amsmath}
\usepackage{amssymb}
\usepackage{xcolor}
\usepackage[range-units = single,range-phrase = \text{--},separate-uncertainty = true,detect-all]{siunitx}
	\DeclareSIUnit\linepair{lp}
	\DeclareSIUnit\pixels{px}
\usepackage{soul}
\setstcolor{red}
\usepackage{cancel}

\definecolor{bleudefrance}{rgb}{0.19, 0.55, 0.91}

\newcommand{\inm}{\textrm{in}}
\newcommand{\outm}{\textrm{out}}

\newcommand{\R}{\mathbf{R}}

\begin{document}

\title{Self-Portrait of the Focusing Process in Speckle: \\ I. Spatio-Temporal Imaging of Wave Packets in Complex Media}

\author{Elsa Giraudat}
\affiliation{Institut Langevin, ESPCI Paris, PSL University, CNRS, 75005 Paris, France}
\author{Flavien Bureau}
\affiliation{Institut Langevin, ESPCI Paris, PSL University, CNRS, 75005 Paris, France}
\author{William Lambert}
\affiliation{SuperSonic Imagine, Aix-en-Provence, France}
\author{Mathias~Fink}
\affiliation{Institut Langevin, ESPCI Paris, PSL University, CNRS, 75005 Paris, France}
\author{Alexandre Aubry$^*$}
\affiliation{Institut Langevin, ESPCI Paris, PSL University, CNRS, 75005 Paris, France}

\date{\today}
\begin{abstract}
    \textbf{This is the first article in a series of three dealing with the exploitation of speckle for imaging purposes. Speckle is the complex interference wave-field produced by a random distribution of un-resolved scatterers. In this paper, we show how these scatterers can be used as virtual microphones to monitor the spatio-temporal propagation of a wave-packet inside the medium. To do so, the concept of matrix imaging is particularly useful. It consists in decoupling the location of the transmitted and received focal spots in a standard beamforming process. By scanning the wave-field with the output focal spot that then acts as a virtual transducer, one can image the spatio-temporal evolution of the wave-packet inside the medium. This unique observable will allow us to highlight the imperfections of the focusing process, in particular the defocus and reverberations induced by a strong aberrating layer. As a proof-of-concept, we will consider ultrasound experiments on tissue-mimicking phantoms. In the next two papers, we will show how this observable can be leveraged to compensate for these phenomena that hamper wave focusing and imaging in all fields of wave physics. Our method is indeed broadly applicable to different types of waves beyond ultrasound for which multi-element technology allows a reflection matrix to be measured.}
\end{abstract}
\maketitle

A speckle pattern is the result of a random interference between many waves. In optics, it is commonly observed at the output of a scattering medium illuminated by coherent light~\cite{Goodman2007}. In medical ultrasound, the speckle pattern arises in the image itself~\cite{Wagner1983}. It is the result of a random interference between echoes induced by a random distribution of un-resolved scatterers. A speckle pattern is characterized by its local short-range correlations both in the spatial and time domain~\cite{Mosk2012}. However, strong correlations can exist between the incident wave-field and transmitted/reflected speckle patterns~\cite{feng_correlations_1988,freund_memory_1988,shahjahan_random_2014,osnabrugge_generalized_2017}. This memory effect has been exploited both in transmission, for imaging through strongly scattering media~\cite{Bertolotti2012,katz_non-invasive_2014} or, in reflection, for adaptive focusing~\cite{kang_high-resolution_2017,lambert_distortion_2020}. In this paper, we will show how speckle can be leveraged to image the propagation of the incident wave-field inside the medium. 

To that aim, we will go beyond confocal imaging which consists in focusing onto a given point inside the medium and only collect echoes coming from the same point. This process can be done physically via a pinhole placed before the detector in optical microscopy or numerically by applying time delays to the received signals in ultrasound. Here, we will decouple the focusing process at input and output in order to scan the incident wave-field with the output focal spot. Assuming speckle is everywhere inside the medium, we will be able to make a time movie of the incident wave inside the medium by exploiting the temporal resolution of our measurement.        

To do so, we will exploit the concept of matrix imaging~\cite{lambert_reflection_2020}. Experimentally, it relies on the measurement of the reflection matrix that contains the impulse response between a set of sensors that are placed on the same side of the medium. Once this matrix is measured, the response of the medium is known for any incident wave-field. Due to the linearity of the wave equation, one can apply an independent beamforming process at input and output in order to investigate the reflection matrix in the focused basis~\cite{lambert_reflection_2020,velichko_quantification_2020}. Each line of this focused reflection matrix contains the incident wave-field intended to focus at $\mathbf{r}_{\textrm{in}}$ recorded by a virtual transducer at $\mathbf{r}_\textrm{out}$. This virtual transducer is synthesized by the focused beamforming process applied at output. This matrix has been, so far, only investigated at the ballistic time to probe the lateral extent of the incident wave-field and evaluate the focusing quality at this specific time~\cite{lambert_reflection_2020,lambert_ultrasound_2022a}. Yet, wave velocity fluctuations can imply a mismatch between the focusing plane dictated by the spatial focusing process and the coherence or isochronous volume controlled by the echo time. Moreover, in presence of reverberations, each scatterer of the medium will not only generate one single echo but several echoes related to each multiply scattered path that can follow the incident and reflected wave-fronts during their travel from the medium surface and the focusing point.  The goal of this study is therefore to exploit the time-dependence of the focused reflection matrix to probe these phenomena. 

Two ultrasound data sets will be used to support our demonstration. The first one consists in the reflection matrix associated with an ultrasonic probe place in contact of a tissue-mimicking phantom whose background sound velocity is known. It will be used as a reference experiment in which the level of aberrations can be tuned by changing the wave velocity model in the beamforming process. The second data set corresponds to the same configuration except that a Plexiglas plate is placed between the probe and the tissue-mimicking phantom in order to mimic a reverberating layer. On the one hand, the bright reflectors contained by the phantom will be used to assess the impact of aberrations and reverberations on the ultrasound image. On the other hand, we will show how ultrasound speckle can be exploited to make a self-portrait of the focusing process, seen through the prism of the wave velocity model. To do so, a coherent virtual reflector will be synthesized from different realizations of input focal spots. In practice, this will be done by means of an an iterative phase reversal algorithm that will be applied to the focused reflection matrix expressed in a de-scanned basis. However, contrary to a previous study~\cite{bureau_three-dimensional_2023}, this process is here applied not only at a central frequency and at the expected ballistic time, but over the whole frequency bandwidth. The result is a movie of the coherent wave-packet that survives to disorder and that highlights: (i) the time shift of the coherence plane and defocus of the wave with respect to the expected ballistic time and focal plane, respectively, when a mismatch exists between the speed-of-sound distribution inside the propagation medium and the wave velocity model used by the beamforming process; (ii) the temporal dispersion of the wave-packets induced by multiple reflections induced by a reverberating layer placed between the probe and the focusing plane. At last, we will discuss how this unique observable can be used to design a path towards: (i) a compensation of axial aberrations (including defocus and reverberations) in the speckle regime and; (ii) a quantitative measurement of the speed-of-sound in reflection. These perspectives will be the object of the two other papers of the series~\cite{Bureau2024,Giraudat2025b}.
\begin{figure*}[t!]
  \includegraphics[width=\linewidth]{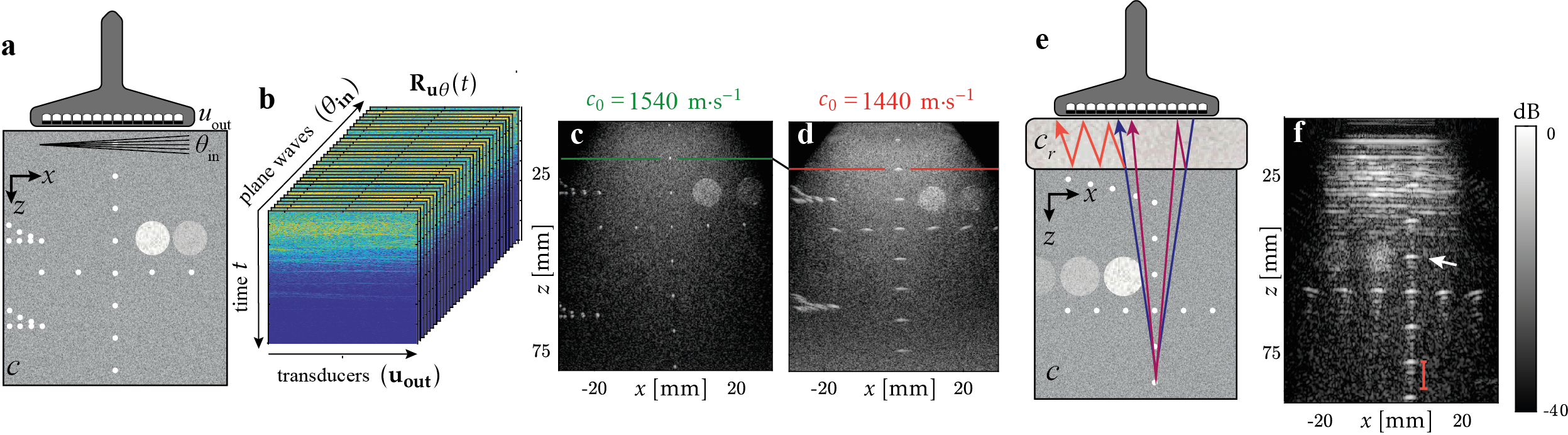}
    \caption{\textbf{Experimental configurations and procedure.} \textbf{a}, Experimental set up:  An ultrasonic linear probe is placed in contact of a tissue-mimicking phantom composed of a random distribution of sub-resolved diffusers schematized by the gray background texture, two hyper-echoic cylinders, and a set of bright nylon rods. A set of plane waves ($\bm{\theta}_{\textrm{in}}$) is emitted by the probe and the time-dependent back-scattered wave-field $R(\mathbf{u}_{\textrm{out}},\bm{\theta}_{\textrm{in}},t)$ recorded by each transducer ($\mathbf{u}_{\textrm{out}}$) is stored in the so-called reflection matrix $\mathbf{R}_{\mathbf{u}\bm{\theta}}(t)=[R(\mathbf{u}_{\textrm{out}},\bm{\theta}_{\textrm{in}},t)]$ sketch in panel $\mathbf{b}$. $\mathbf{c}$, $\mathbf{d}$  Ultrasound image deduced from a beamforming process applied to $\mathbf{R}_{\mathbf{u}\bm{\theta}}(t)$ using $c_0=1540$  and 1440 m.s$^{-1}$, respectively. \textbf{e}, Same experimental configuration as in panel \textbf{a} except that a reverberating layer is placed between the probe and the tissue-mimicking phantom. \textbf{f}, Corresponding ultrasound image ($c_0=1540$ m.s$^{-1}$).}    
    \label{fig1}
\end{figure*}

\section{Time focused reflection matrix}

\subsection{Reflection matrix {acquisition}} 

Ultrasound matrix imaging (UMI) begins with the acquisition of the reflection matrix using an ultrasound array of transducers (Figs.~\ref{fig1}a,b and Tab.~\ref{Table_SondeLin_ExpPhantom}). The sample under study is a tissue mimicking phantom with a speed-of-sound $c=1542 \pm 10$ m/s. It is composed of a random distribution of unresolved scatterers which generate an ultrasonic speckle characteristic of human tissue (Fig.~\ref{fig1}c). It also contains a set of nylon rods. Those bright scatterers can be used to evaluate locally the imaging point spread function (PSF) of the ultrasound device. The reflection matrix is captured by sending a series of plane waves into the medium. Each plane wave is identified with its angle of incidence ${\theta}_{\textrm{in}}$ (Fig.~\ref{fig1}a). For each illumination, the reflected waves are recorded by the transducers of the probe, each element being identified by its lateral position $u_{\textrm{out}}$ (Fig.~\ref{fig1}b). The recorded wave-fronts are noted $R(u_{\textrm{out}},{\theta}_{\textrm{in}},t)$, with $t$ the echo time. They are stored in a reflection matrix ${\mathbf{R}_{{u}{\theta}}(t)=[R(u_\outm,\theta_{\textrm{in}},t)]}$.
\begin{table}[h!tb]
   \centering
      \begin{tabular}{|c|l|l|l|} 
      \hline
      \multicolumn{3}{|c|}{\textbf{parameter}} & \textbf{value} \\
      \hline
      \hline
      \multirow{8}*{\textbf{probe}} & \multicolumn{2}{l|}{type} & linear \\
       & \multirow{2}*{number of transducers} & transmission & $N_u^{\mathrm{(Tx)}} = 256$ \\
       &   & Reception & $N_u^{\mathrm{(Rx)}} = 128$ \\
       & \multicolumn{2}{l|}{inter-element distance}  & $\delta u = 0.2$~\unit{\milli\metre} $\sim  \lambda$\\
       & \multirow{2}*{aperture} & transmission & $\Delta u^{\mathrm{(Tx)}} = 51.2$~\unit{\milli\metre} \\
       &   & reception & $\Delta u^{\mathrm{(Rx)}} = 25.6$~\unit{\milli\metre}\\
       & \multicolumn{2}{l|}{central frequency}  & $f_c = 7.5$~\unit{\mega\hertz} \\
       & \multicolumn{2}{l|}{bandwidth}  & $\Delta f = \left[ 2 - 10\right]$~\unit{\mega\hertz} \\
      \hline
      \multirow{8}*{\textbf{acquisition}} & \multicolumn{2}{l|}{imaging platform} & Aixplorer\textregistered, Supersonic Imagine\\
        & \multicolumn{2}{l|}{speed-of-sound model} & $c_0 = 1540$~\unit{\metre\per\second }\\  
       & \multirow{3}*{plane waves} & maximum angle & $\theta_{\textrm{max}}= \ang{40}$  \\
       &  & angular sampling & $\delta\theta_{\textrm{in}} = \ang{1}$  \\
       &  & number & $N_{\textrm{in}} = 81$\\
       & \multicolumn{2}{l|}{emitted signal} &  two half-period pulse at $f_c$   \\
       & \multicolumn{2}{l|}{sampling frequency} & $f_s = 30$~\unit{\mega\hertz}\\
        & \multicolumn{2}{l|}{recording time} & $\Delta t = 137$~\unit{\micro\second}\\
       \hline
   \end{tabular}
   \caption{{Probe and acquisition sequence used for recording the canonical reflection matrix $\R_{u\theta}(t)$ in the experiments.}}
   \label{Table_SondeLin_ExpPhantom}
\end{table}

\subsection{Confocal imaging}

\begin{figure*}[t!]
  \includegraphics[width=1\linewidth]{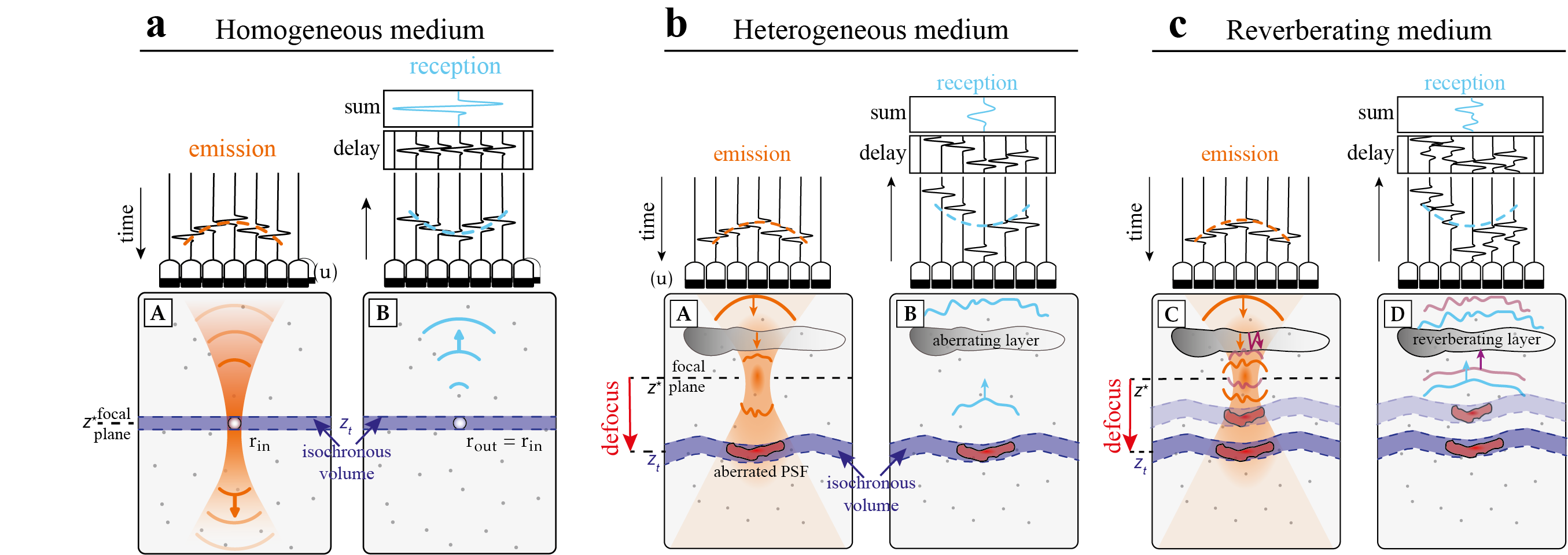}
    \caption{\textbf{Principle of ultrasound imaging.} \textbf{a}, Case of a medium of homogeneous speed-of-sound. Left: The excitation of each transducer by a time-delayed pulse generates a cylindrical wave-front focusing at a given point $\mathbf{r}_{\textrm{in}} = (x_{\textrm{in}}, z_{\textrm{in}})$. Right: In reception mode,  the echoes back-scattered by reflactors in the vicinity of the same point are time-shifted using the same delay law as that applied in transmit. This allows the signals induced by a single scattering event at $\mathbf{r}_{\textrm{out}} =\mathbf{r}_{\textrm{in}} $ to be constructively summed. The pixel of the ultrasound image corresponds to the resulting signal measured
at the expected ballistic time $t = 2z_{\textrm{in}}/c_0$. The set of points generating the backscattered echoes at  time $t$ defines the isochronous volume. When the wave velocity model and the medium speed-of-sound coincide, the focusing plane of the incident wave-front belongs to the isochronous volume. \textbf{b}, For a medium with an heterogeneous speed-of-sound distribution, this coincidence is no longer checked: The focusing plane and the isochronous volume are axially shifted in reverse directions. The delay-and-sum beamforming process that assumes an homogeneous wave velocity $c_0$ is hampered by a defocus. \textbf{c}, For a reverberating medium, the situation is even worse since each multiply-reflected wave-front is associated with a different isochronous volume and a distinct focusing plane. Reverberation phenomena undergone by the wave at input and output gives rise to a series of echoes for each scatterer on the beamformed image.}    
    \label{fig2}
\end{figure*}

The first post-processing step is to build a confocal image $\mathcal{I}$ from the recorded reflection matrix. To do so, a delay-and-sum beamforming process is applied to the coefficients of ${\mathbf{R}_{{u}{\theta}}}(t)$. Physically equivalent to a confocal focusing process (Fig.~\ref{fig2}a), this procedure writes mathematically as follows:
\begin{align}
\label{confocal}
\mathcal{I}(x,z_0=c_0t/2)&=  \sum_{\theta_\inm}\sum_{u_\outm} A(u_\outm,\theta_\textrm{in},x,t,c_0)  \\  & R(u_\outm,\theta_\textrm{in},\tau_{\textrm{out}}(u_\outm,x,t,c_0)
+\tau_{\textrm{in}}(\theta_\textrm{in},x,t,c_0)). \nonumber
\end{align}
$c_0$ is the wave velocity model considered in the beamforming process. $\tau_{\textrm{in}}$ is the time-of-flight expected for the incident plane wave to reach the target point of coordinates $(x,z_0)$. $\tau_{\textrm{out}}$ is the time-of-flight expected for the reflected wave to travel from the same target point to each transducer. $A$ is a normalization and apodization factor that limits the extent of the
receive synthetic aperture. $z_0=c_0t/2$ is the expected position of the isochronous volume, which is defined as the ensemble of points that contribute to the ultrasound signal at time $t$. When the wave velocity model $c_0$ matches with the medium speed-of-sound, the position $z_t$ of the isochronous volume matches with the focusing plane $z^{\star}$. The ultrasound image is then a satisfactory image of the medium reflectivity (Fig.~\ref{fig1}c). In presence of an aberrating layer associated with a speed-of-sound $c$ differing from $c_0$, the beamformed image is drastically affected by the mismatch between the isochronous volume and the focusing plane (Fig.~\ref{fig2}b). This non-coincidence results in axial aberrations on the ultrasound image (Fig.~\ref{fig1}d) that manifest as: (\textit{i}) an axial shift of the scatterers with respect to their true depth; (\textit{ii}) a degradation of the transverse resolution. In presence of reverberations (Fig.~\ref{fig1}e), each multiply-reflected wave is associated with a distinct focusing depth and a shifted isochronous volume (Fig.~\ref{fig2}c). Those reverberations induce multiple ghost images of each bright scatterer on the ultrasound image (Fig.~\ref{fig1}f), thereby highlighting the spatio-temporal spreading of the imaging PSF. However, speckle statistics does not seem affected by aberrations and reverberations in Figs.~\ref{fig1}d and f. Yet, speckle also carries the spatio-temporal distortions undergone by the incident and reflected wave-fields upstream. In the following, we will show how to reveal this hidden information with the help of matrix imaging. 

\subsection*{Time Focused Reflection Matrix} 

\begin{figure*}[t!]
  \includegraphics[width=1\linewidth]{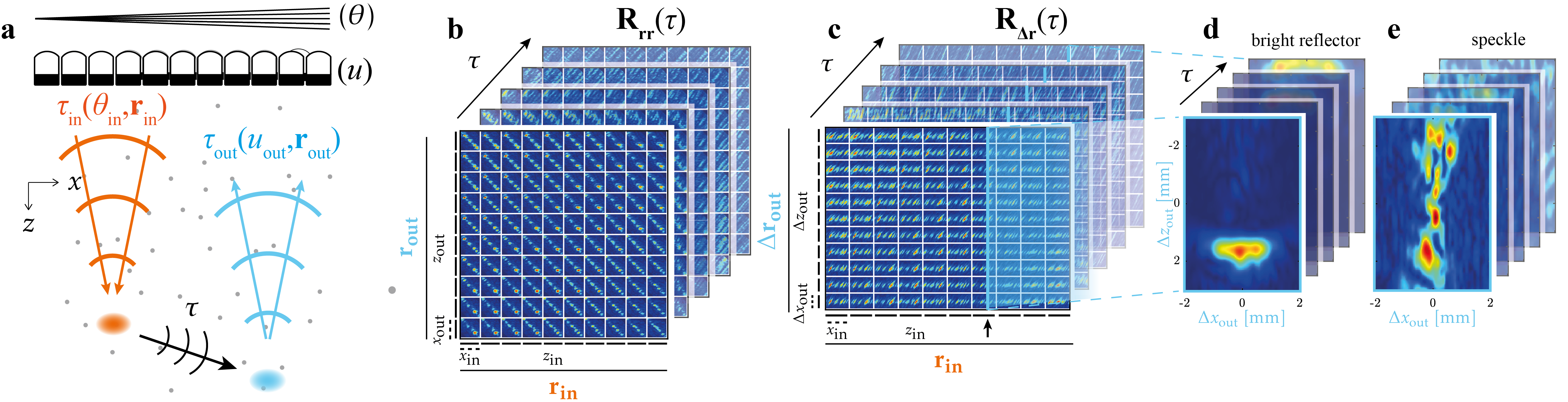}
    \caption{\textbf{Time-focused reflection matrix.} \textbf{a} The matrix $\mathbf{R_{rr}}(\tau)$ is constructed in post-processing by decoupling the position of the emission focusing points $\mathbf{r}_{\textrm{in}} = (x_{\textrm{in}},z_{\textrm{in}})$ and reception focal points $\mathbf{r}_{\textrm{out}} = (x_{\textrm{out}},z_{\textrm{out}})$, thus forming a set of virtual sources and receivers inside the medium. The temporal responses between each virtual source $\mathbf{r}_{\textrm{in}} $ and each virtual receiver $\mathbf{r}_{\textrm{out}} $ are obtained by introducing an additional variable time $\tau$, corresponding to a shift from the expected ballistic time $t = \left(z_{\textrm{in}}+ z_{\textrm{out}} \right)/c_0$. This time delay $\tau$ allows the wave to propagate virtually in the medium from $\mathbf{r}_{\textrm{in}} $ to $\mathbf{r}_{\textrm{out}} $. \textbf{b} Time focused reflection matrix $\mathbf{R_{rr}}(\tau)$ displayed at different time lapses $\tau$ in the ultrasound phantom (Fig.~\ref{fig1}a). \textbf{c} Same matrix $\mathbf{R_{\Delta r}}(\tau)$ expressed in the de-scanned basis. \textbf{d} Example of a reflected wave-fields reshaped in de-scanned coordinates  $\left(\Delta x, \Delta z \right)$ extracted  from one column of the de-scanned matrix $\mathbf{R}_{\bm{\Delta} \mathbf{r}}(\tau)$ designated by a black arrow in panel c and associated with a given point $\mathbf{r}_{\textrm{in}}$ corresponding to a bright point-like scatterer. \textbf{e} Same as in panel d but in speckle. }    
    \label{fig3}
\end{figure*}

Ultrasound matrix imaging can show the degradation of the focusing process by decoupling the input and output focusing points, $\mathbf{r}_{\textrm{in}}$ and $\mathbf{r}_{\textrm{out}}$. This process is achieved by projecting the reflection matrix in a focused basis. This operation can be performed through a simple matrix product in the frequency domain~\cite{lambert_reflection_2020}, or in the time domain~\cite{bureau_three-dimensional_2023}. In the latter option, it consists in a delay-and-sum beamforming process generalized to the case of different focal points in transmission and reception modes:
\begin{equation}
   R\left(\mathbf{r}_{\textrm{out}},\mathbf{r}_{\textrm{in}},\tau \right) = \sum_{\theta_{\textrm{in}}} \sum_{u_{\textrm{out}}} A\left({u_{\textrm{out}},\mathbf{r}_{\textrm{out}}},{\theta_{\textrm{in}},\mathbf{r}_{\textrm{in}}}\right) R\left(u_{\textrm{out}},\theta_{\textrm{in}},\tau_{\textrm{out}}(u_\outm,x,t,c_0)
+\tau_{\textrm{in}}(\theta_\textrm{in},x,t,c_0)+\tau \right). 
   \label{chp3_TimeDAS}
\end{equation}
Contrary to previous works, the resulting focused reflection matrix, $\mathbf{R_{rr}}(\tau)=[R(\mathbf{r}_{\textrm{out}},\mathbf{r}_{\textrm{in}},\tau)]$, is here expressed as a function of a time lapse $\tau$ with respect to the expected ballistic time. An example of focused reflection matrix $\mathbf{R_{rr}}(\tau)$ is displayed at different lapse time in Fig.~\ref{fig3}b. Given the 2D feature of points $\mathbf{r}_{\textrm{in}}$ and $\mathbf{r}_{\textrm{out}}$, the full reflection matrix is in five space-time dimensions.  In previous works~\cite{lambert_reflection_2020,lambert_ultrasound_2022a}, the focused reflection matrix was only considered at the ballistic time $\tau = 0$, which limited us to a study of transverse aberrations induced by the variations of the speed-of-sound with respect to its mean value. Yet, as illustrated by Fig.~\ref{fig1}, axial aberrations can be far from negligible especially in presence of reverberations. In the present paper, this supplementary degree of freedom will be exploited to probe the time-dependence of the focusing process.

The focused reflection matrix can be physically interpreted as the inter-element response matrix associated with virtual transducers located in the medium~\cite{lambert_reflection_2020}. At emission, the digital focusing of energy at a point $\mathbf{r}_{\textrm{in}}$ can be interpreted as the synthesis of a virtual source at that point. At reception, the focusing point $\mathbf{r}_{\textrm{in}}$ can similarly be seen as a virtual detector, formed by digitally selecting the back-scattered echoes originating from that point.  It should be noted that the concept of virtual transducers is purely didactic. Of course, these virtual transducers do not act as true active sources of energy. Furthermore, they are highly directional: downward for virtual sources, and upward for virtual receivers~\cite{lambert_ultrasound_2022a}. 

The wave induced by a virtual source at $\mathbf{r}_{\textrm{in}}$ can be measured by scanning the virtual receiver $\mathbf{r}_{\textrm{out}}$ around it. To visualize the propagation of this wave across the medium, the focused reflection matrix can be expressed in a de-scanned frame using the following change of variables:  
\begin{equation}
    R(\Delta \mathbf{r},\mathbf{r}_{\textrm{in}},\tau)=R(\mathbf{r}_{\textrm{out}},\mathbf{r}_{\textrm{in}},\tau)
    \end{equation}
with $\Delta \mathbf{\mathbf{r}}=\mathbf{r}_{\textrm{out}}-\mathbf{r}_{\textrm{in}}$. The resulting de-scanned matrix, $\mathbf{R_{\Delta r}}(\tau)=[ R(\Delta \mathbf{r},\mathbf{r}_{\textrm{in}},\tau)]$, is displayed in Fig.~\ref{fig3}c for the phantom experiment sketched in Fig.~\ref{fig3}a. 
Although it displays a 5D structure as the canonical matrix $\mathbf{R_{r r}}(\tau)$, the de-scanned coordinates $\Delta x$ and $\Delta z$ can display less sampling points than the coordinates $x$ and $z$. It is therefore a convenient basis for storing the ultrasound data. 

Fig.~\ref{fig3}d explains how the information contained in the-scanned matrix will be displayed in the following. A point $\mathbf{r}_{\textrm{in}}$ marked by a black arrow (Fig.~\ref{fig3}c) is chosen for illustration purposes. The corresponding column ($\Delta \mathbf{r}$ ) is extracted and reshaped in cartesian coordinates $\left(\Delta x,\Delta z \right)$, which allows us to visualize the spatio-temporal responses between the virtual source created at this point and all the surrounding virtual receivers. Fig.~\ref{fig3}d and e show the amplitude of the field obtained for a point $\mathbf{r}_{\textrm{in}}$ corresponding to a bright point-like scatterer or in the speckle (right). In the latter case, the field is modulated by the random reflectivity of the scatterers located at points $\mathbf{r}_{\textrm{in}}$ and  $\mathbf{r}_{\textrm{out}}$ used to probe the field.  We will see later that we can overcome the random fluctuations in reflectivity to recover the wave that would be associated with a coherent source located at $\mathbf{r}_{\textrm{in}}$, even in speckle regime.

We will now investigate in greater detail the information contained in the time-focused reflection matrix $\mathbf{R_{\Delta r}}(\tau)$ in the case of each of the experiments introduced above (Fig.~\ref{fig1}). In particular, we will show how this matrix can be used to probe the spatio-temporal focusing process in the medium. To do this, we proceed in two stages. First, we consider the case where the virtual source  $\mathbf{r}_{\textrm{in}}$ coincides with a bright scatterer in the medium . We use a set of virtual reflectors $\mathbf{r}_{\textrm{out}}$  located around  $\mathbf{r}_{\textrm{in}}$ to probe the reflected field from this source. In this case, the bright scatterer acts as an acoustic guide star located in the medium. However, such a strong reflector is not often available in vivo and not everywhere in any case. This is why, in a second step, we will consider the more complex case of a virtual source located in speckle. In this common regime, we will see that it is also possible to probe the focusing process. To do this, we will use an iterative phase reversal algorithm to synthesize a coherent virtual reflector from a set of incoherent sources $\mathbf{r}_{\textrm{in}}$ generated in the speckle.

\section{Self-portrait of a wave reflected by a bright scatterer} 

In ultrasound imaging, isolated bright scatterers can be used as guide stars for adaptive focusing. Here we show how the wave-field reflected by such guide star can be monitored thanks to virtual receivers synthesized in speckle. The movie of this wave-field bears the signature of the axial distortions and reverberations undergone by the wave accross the medium.

\subsection{Correct propagation model}

\begin{figure*}[t!]
  \includegraphics[width=0.8\linewidth]{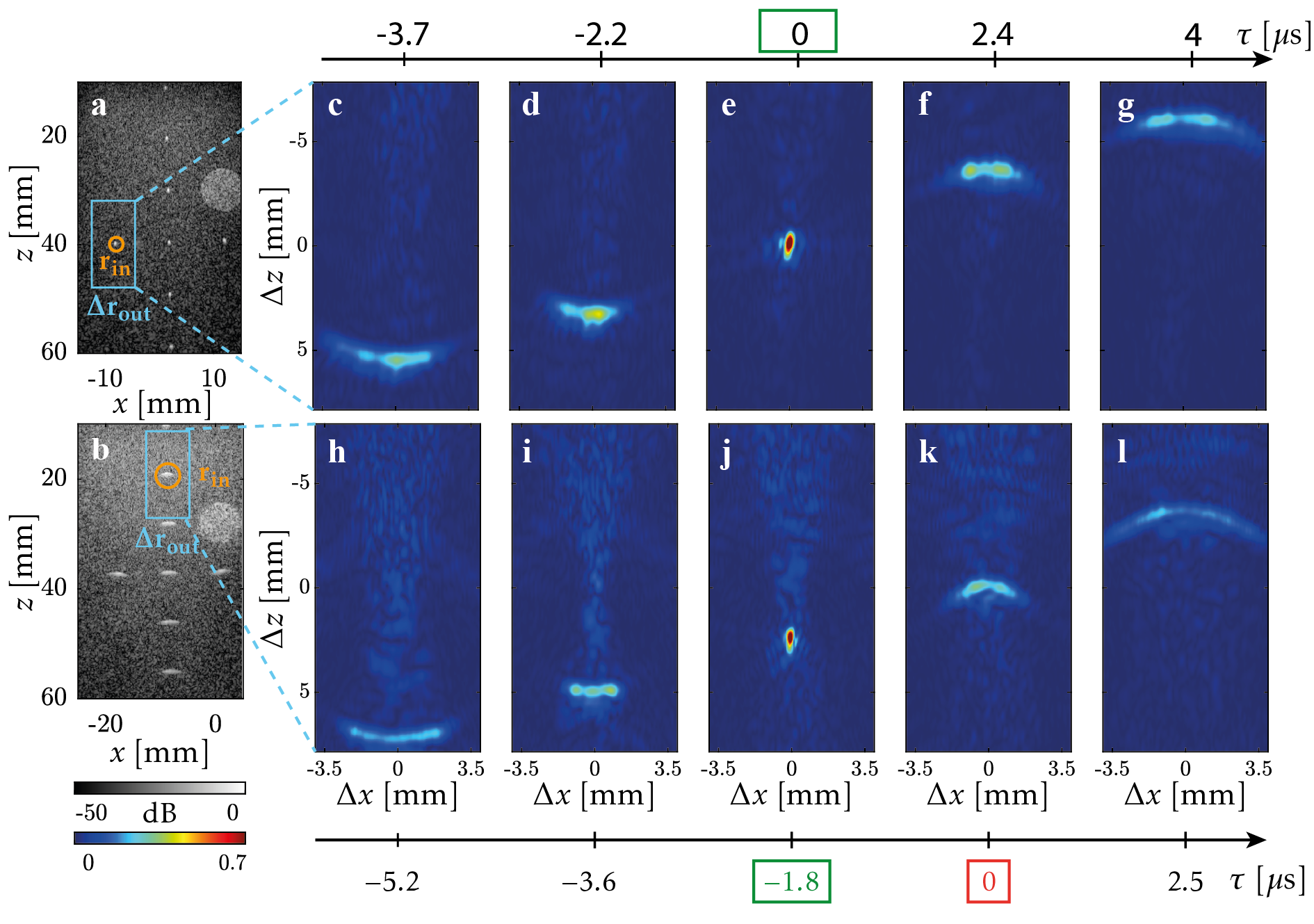}
    \caption{\textbf{Coherent wave-packet associated with a bright scatterer inside an homogeneous phantom.}  \textbf{a},\textbf{b} Ultrasound image of the phantom beamformed at the phantom speed-of-sound ($c_0=1540$ m.s$^{-1}$) and at a wrong speed-of-sound ($c_0=1440$ m.s$^{-1}$), respectively: The position of the virtual source $\mathbf{r}_{\textrm{in}}$ corresponds to a bright scatterer indicated by an orange cross and the set of virtual receivers $\mathbf{r}_{\textrm{out}}$ surrounding it is delimited by a blue rectangle. 
\textbf{c}-\textbf{g}, Normalized amplitude of the wave-field $R(\Delta \mathbf{r},\mathbf{r}_{\textrm{in}},\tau)$ produced  by the virtual source $\mathbf{r}_{\textrm{in}}$ at the phantom speed-of-sound (a). \textbf{h}-\textbf{l} Same as in (c-g) but for a wrong speed-of-sound (b). In panels (c-l), each focused wave-field is displayed in de-scanned coordinates for various lapse times $\tau$. The propagation of the coherent wave-packet shown in panels (c-g) and (h-l) are also displayed in Supplementary Movies 1 and 2.}
    \label{fig4}
\end{figure*}

First, let us consider the ``ideal'' case of the ultrasonic phantom experiment with a constant
wave speed $c_0 = 1542$ m.s$^{-1}$ (Fig.~\ref{fig1}a). We are considering a virtual source $\mathbf{r}_{\textrm{in}}$ coinciding with the bright scatterer indicated by an orange cross in Fig.~\ref{fig4}a, as well as a set of virtual receivers distributed around this source,
as delimited by the blue rectangle in Fig.~\ref{fig4}a. The column of the matrix corresponding to this point $\mathbf{r}_{\textrm{in}}$ gives access to the propagation film of the refocused wave associated with this virtual source. This field is
represented in the de-scanned coordinate system $(\Delta x, \Delta z)$ for several lapse times $\tau$ (Figs.~\ref{fig4}b-f), as explained in Fig.~\ref{fig3}d. In this configuration, the wave reflected by the nylon rod clear predominates over the echoes produced by surrounding sub-resolved scatterers. We can therefore observe a coherent wave propagating in the medium, from deep $z$ towards the probe. The wave is converging towards $\mathbf{r}_{\textrm{in}}$ at negative times $\tau$, then focuses at time $\tau = 0$ at $(\Delta x, \Delta z)=(0,0)$, and finally diverges
at positive times $\tau$. The anti-causal component of this coherent wave (Figs.~\ref{fig4}c,d) corresponds to the pressure wave-field that would produce the virtual source at $\mathbf{r}_{\textrm{in}}$ (Figs.~\ref{fig4}e) in a transmission configuration. The causal part of this coherent wave corresponds to the wave-field induced by this virtual source (Figs.~\ref{fig4}f,g).  The time focused reflection matrix thus provides access to the propagation film of the reflected wave when a bright scatterer is positioned at $\mathbf{r}_{\textrm{in}}$. This statement is valid when the velocity model corresponds to the actual speed-of-sound distribution inside the propagation medium. However, in practice, it is rare to have prior knowledge of the wave velocity in the medium, so the model is generally only a more or less refined approximation of the actual velocity.

\subsection{Probing axial aberrations}

Let us now examine how the time focused reflection matrix can be also used to highlight a focusing process degraded by an erroneous velocity model. The mismatch between the $c_0$ model and the actual velocity distribution in the medium implies a shift between the isochronous volume and the spatial focal plane (Fig.~\ref{fig2}b). This results in transverse spreading of the focal spots and a degradation of the resolution and contrast of ultrasound images (Fig.~\ref{fig1}d). In this section, we will see how this shift between the isochronous volume and the focal plane manifests itself on the time-focused reflection matrix. In particular, we will show how this observable is a suitable tool for probing this shift and, ultimately, for resolving it.
 
As in Fig.~\ref{fig1}d, the case of an incorrect velocity model is artificially reproduced by deliberately constructing the focused reflection matrix with a velocity model $c_0 = 1440$ m.s$^{-1}$ that differs from the speed-of-sound $c = 1542$ m.s$^{-1}$ in the phantom. As in the previous section, we extract the focused wave-field corresponding to a virtual source $\mathbf{r}_{\textrm{in}}$ coinciding with a bright scatterer (Fig.~\ref{fig4}c). As for a correct propagation model (Fig.~\ref{fig4}b), the propagation film reveals a coherent wave propagating towards the probe at a wave speed $c_0$ (Fig.~\ref{fig4}d). However, this movie bears the mark of the mismatch between the velocity model and the true speed-of-sound. Wave focusing is observed neither at the desired location ($\Delta z^{\star} =2.5$ mm) nor at the expected ballistic time  ($\tau^{\star} =- 1.8 $ $\mu$s).  At the ballistic time $\tau=0$, the wave associated with the strong scatterer is therefore out-of-focus, which explains the distorted shape of the target on the confocal image (Fig.~\ref{fig4}c).

To explain quantitatively how this shift in focusing time and depth is related to the phase velocity mismatch, we can express the de-scan reflection matrix associated with a point-like scatterer of reflectivity $\gamma_s$ at position $(x_{s}, z_s)$. To do so, the focused reflection matrix will be expressed in the frequency domain:  $ \overline{\mathbf{R}}_{\mathbf{rr}}(f)=\int d\tau {\mathbf{R}}_{\mathbf{rr}}(\tau) \exp \left ( -j 2 \pi f \tau \right) $. The coefficients of the monochromatic focused reflection matrices can be theoretically expressed as follows~\cite{lambert_reflection_2020,lambert_ultrasound_2022a}:  
\begin{equation}
\label{Rxx}
   \overline{R}(x_{\textrm{out}},x_{\textrm{in}},z_{\textrm{out}},z_{\textrm{in}},f) =\gamma_s \overline{B}(f) H_{\textrm{out}}(x_s,z_s,x_{\textrm{out}},z_{\textrm{out}},f)  H_{\textrm{in}}(x_{\textrm{in}},z_s,x_{\textrm{in}},z_{\textrm{in}},f),
\end {equation}
where $\overline{B}(f)$ accounts for the probe bandwidth. {$H_{\textrm{in/out}}(x_s,z_s,x_{\textrm{in/out}},z_{\textrm{in/out}},f)$} corresponds to the monochromatic PSF, i.e., the spatial distribution of the focal spot amplitude at the scatterer position when trying to focus on the point $\mathbf{r}_{\textrm{in/out}} = (x_{\textrm{in/out}},z_{\textrm{in/out}})$. Under the paraxial approximation, the PSF can be, in first approximation, decomposed as a product of a frequency-invariant enveloppe and a depth-oscillating phase term   (see Appendix~\ref{A}):
\begin {equation}
   H_{\textrm{in/out}}\left(x,z,x_{\textrm{in/out}},z_{\textrm{in/out}},f \right) = \underbrace{H^{(0)}_{\textrm{in/out}} \left(x-x_{\textrm{in/out}},cz/c_0-z_{\textrm{in/out}},f_c\right)}_{\textrm{transverse focusing}}  \underbrace{e^{i {2\pi f}\left(\frac{z_{\textrm{in/out}}}{c_0} -\frac{z}{c}\right)}}_{\substack{\text{axial propagation}}} \label{ch3_eq_monof_H}
\end {equation}
While the first term in Eq.~\ref{ch3_eq_monof_H} accounts for the transverse focusing process at the focusing depth, 
\begin{equation}
    z^{\star}_{\textrm{in/out}}=c_0z_{\textrm{in/out}}/c,
\end{equation} 
the second term accounts for the axial propagation of the focused wave-field along the $z-$axis. The scatterer depth $z_s$ is controlled by the time-of-flight expected for a scatterer at the targeted depth $z_\textrm{in}$ under a wave velocity model $c_0$:  $z_s=\frac{c}{c_0} z_{\textrm{in}}$.
Using this expression of $z_s$ and injecting  Eq.~\ref{ch3_eq_monof_H} into Eq.~\ref{Rxx} leads to the following expression of the focused $\mathbf{R}-$matrix coefficients in de-scanned coordinates in the time domain,
\begin{equation}
   {R}(\Delta x,\Delta z,z_{\textrm{in}},\tau) \propto \underbrace{H_{\textrm{in}}^{(0)}\left(0,\Delta z^{\star} ,f_c\right)}_{constant} \underbrace{H_{\textrm{out}}^{(0)}\left(\Delta x,\Delta z^{\star}-\Delta z ,f_c\right)}_{\textrm{spatial focusing}} \underbrace{B\left( \tau+ \frac{\Delta z} {c_0}\right)}_{\textrm{time focusing}}.
   \label{ch3_eq_monof_Rdrr2}
\end{equation}
with
\begin{equation}
    \Delta z^{\star}=\left (\frac{c^2}{c_0^2} -1\right )z_{\textrm{in}},
\end{equation}
the focusing depth in the de-scanned frame. The temporal term in Eq.~\ref{ch3_eq_monof_Rdrr2} proves the upward propagation of the wave-front at the supposed speed-of-sound $c_0$. The propagation films thus provide a self-portrait of the focusing process under the prism of the wave velocity model. Despite this bias, the observed focusing time depends on the actual velocity $c$ in the propagation medium:
\begin{equation}
\label{tau_star}
    \tau^\star=-\Delta z^{\star}/c_0=\frac{z_{\textrm{in}}}{c_0}\left (1-\frac{c^2}{c_0^2} \right ).
\end{equation}
 A numerical application using the apparent depth value $z_{\textrm{in}} = 18.9$ mm of the scatterer selected in Fig.~\ref{fig4}c yields values of $\Delta z^{\star} \sim 2.7$ mm and $\tau^{\star} \sim -1.88$ $\mu$m, which are consistent with the experimental values previously extracted from Fig.~\ref{fig4}d.  

The propagation movie of the reflected wave therefore provides a relevant information about the discrepancy between this velocity model and reality. Interestingly, we now show how such an observable can be used to probe reverberations induced by a layer displaying a strong impedance mismatch with respect to its environment.

\subsection{Probing reverberations}

\begin{figure*}[t!]
  \includegraphics[width=1\linewidth]{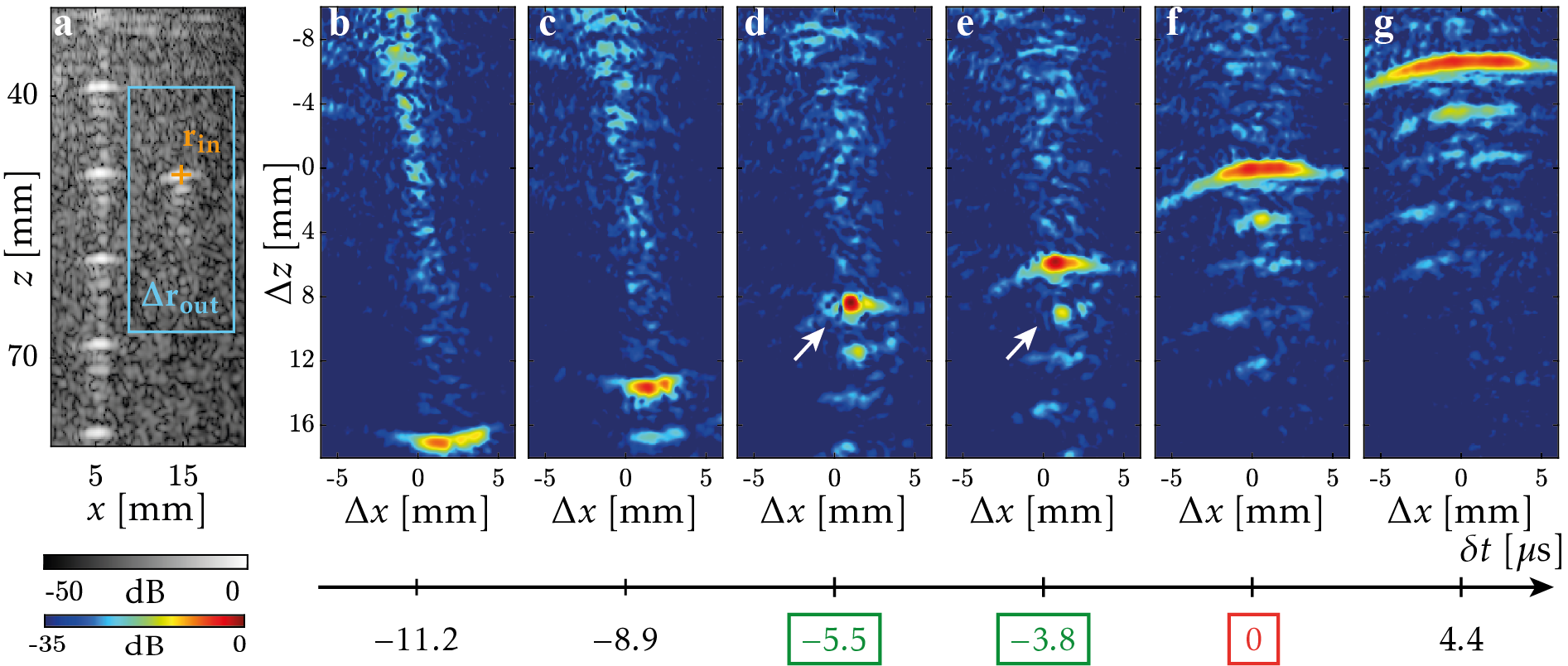}
    \caption{\textbf{Coherent wave-packet associated with a bright scatterer in presence of reverberations.}  \textbf{a}, Ultrasound image of the phantom through the Plexiglas layer (Fig.~\ref{fig1}e): The position of the virtual source $\mathbf{r}_{\textrm{in}}$ corresponds to a bright scatterer indicated by an orange circle and the set of virtual receivers $\mathbf{r}_{\textrm{out}}$ surrounding it is delimited by a blue rectangle.
\textbf{b}-\textbf{f}, Normalized amplitude of the wave-field produced by the virtual source $\mathbf{r}_{\textrm{in}}$ in de-scanned coordinates displayed for various lapse times $\tau$. The propagation of the coherent wave-packet is also displayed in Supplementary Movie 2.}    
    \label{fig5}
\end{figure*}
We now focus on the influence of a reverberating plate on the focusing process. To do so, we consider the experiment with the ultrasonic phantom covered with a Plexiglas plate of thickness $L=6$ mm and speed-of-sound $c_r=2750$ m.s$^{-1}$ (Fig.~\ref{fig1}e). The focused reflection matrix is synthesized assuming an homogeneous velocity model $c_0$ equal to the speed-of-sound in the phantom ($c = 1540$ m.s$^{-1}$). This choice ensures, in first approximation, isoplanicity over the whole phantom. Only the finite size of the probe limits the spatial invariance of the imaging configuration.

\begin{figure}[h!tb]
    \includegraphics[width=1\linewidth]{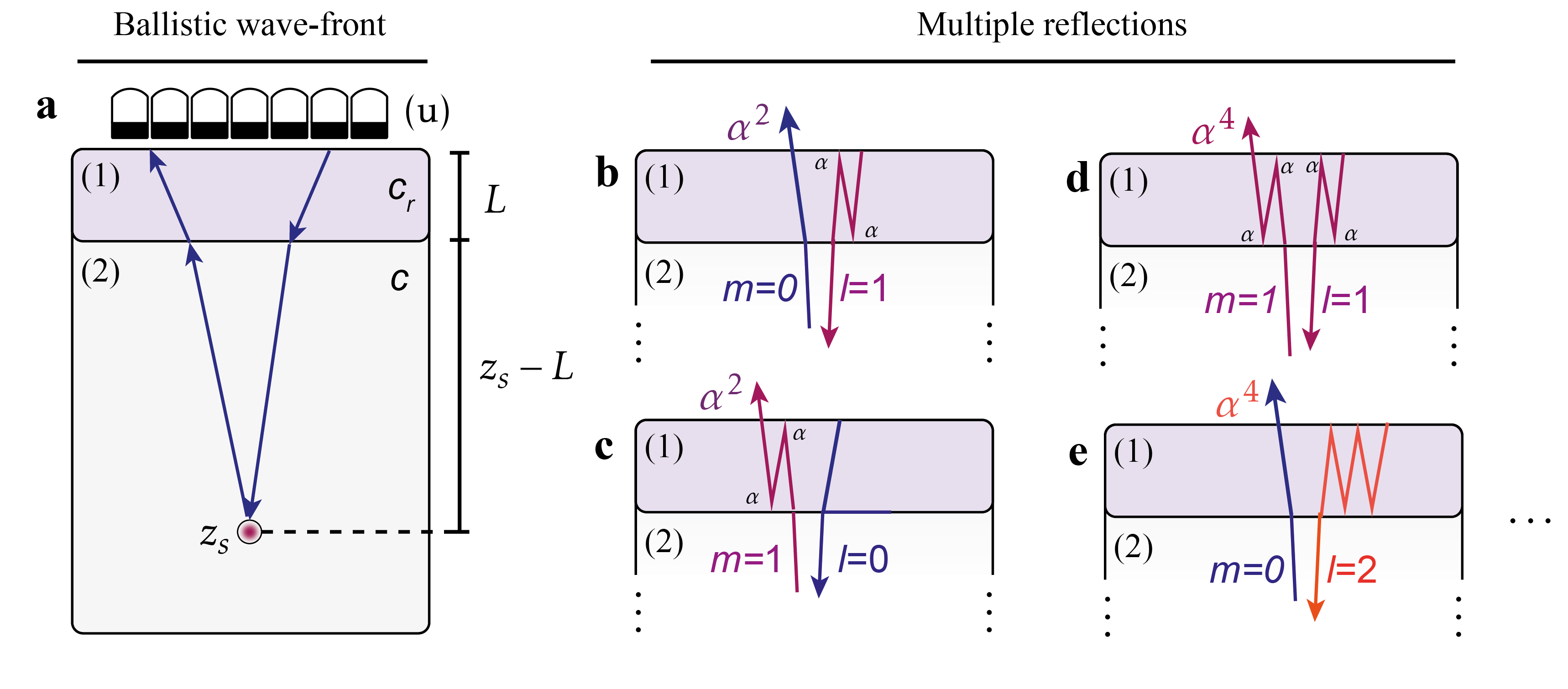}
     \caption{\textbf{Diagram of the different wave propagation paths in a two-layer medium.} \textbf{a} Example of ballistic path trajectory between the probe and a scatterer at $z_s$ through a reverberating layer of thickness $L$. \textbf{b}, Example of first-order reflection path trajectory associated with the reflection matrix component $\mathbf{R}^{(1,0)}$ (Eq.~\ref{decompose}): The incident wave is reflected twice at the interfaces of the reverberating layer. \textbf{c}, Example of first-order reflection path trajectory associated with the component $\mathbf{R}^{(0,1)}$: The up-going wave is reflected twice at the interfaces of the reverberating layer. \textbf{d}, Example of second-order reflection path trajectory associated with the component $\mathbf{R}^{(1,1)}$: The incident and reflected waves are reflected twice at the interfaces of the reverberating layer. \textbf{e}, Example of second-order reflection path trajectory associated with the component $\mathbf{R}^{(2,0)}$: The incident wave is reflected four times at the interfaces of the reverberating layer.}
     \label{fig6}
 \end{figure}

The reflected wave-field associated with the bright scatterer indicated in Fig.~\ref{fig5}a is displayed at different lapse times in Fig.~\ref{fig5}b-g. It clearly highlights the presence of multiple reflections associated with a period of $\sim$2 $\mu$s. Those reverberations \textit{a priori} correspond to waves that have been multiply reflected between the probe and the phantom surface, as shown schematically in Fig.~\ref{fig6}. Furthermore, we can observe in the film that the coherent ballistic wave focuses at a time $\tau =-5.5$ $\mu$s, well before the ballistic time expected by the homogeneous velocity model. This difference between the observed focusing time and the ballistic time can be explained by the large gap between the velocity model and the speed-of-sound distribution in the current experiment. The large speed-of-sound in the Plexiglas layer implies this earlier arrival of the direct echo compared to what would be expected in an homogeneous phantom.
 
To analyze more quantitatively our experimental observation, a theoretical analysis of reverberations is now performed by considering a bi-layered medium consisting in an imaging volume of speed-of-sound $c$ placed behind a reverberating layer of thickness $L$ and wave velocity $c_r$ (Fig.~\ref{fig6}). The reflected wave-field can be decomposed into the set of possible scattering paths that the incident and reflected waves can undertake during their travel between the probe and the isochronous volume (Fig.~\ref{fig6}). Under a matrix formalism, it can be written as follows:
\begin{equation}
\label{decompose}
  \mathbf{R}_{\Delta \mathbf{r}} (\tau) = \sum_{l=0}^\infty \sum_{m=0}^\infty  \alpha^{2(m+l)}  \mathbf{R}^{(l,m)}_{\Delta \mathbf{r}}(\tau) .
\end{equation} 
where $\mathbf{R}^{(l,m)}_{\Delta \mathbf{r}}$ accounts for each component of the reflection matrix associated with $2l$ internal reflections on the incident way and $2m$ internal reflections on the return path. The coefficient $\alpha$ accounts for the internal reflection coefficient in the reverberating plate. This reflection coefficient weights the sum over scattering paths in Eq.~\ref{decompose} and controls the number of reverberations we can observe in practice (Fig.~\ref{fig5}).  

In presence of  bright scatterer at $(x_s,z_s)$, each component of the focused reflection matrix (Eq.~\ref{decompose}) can be expressed as follows:
\begin{equation}
\label{Rbright0}
\overline{\mathbf{R}}^{(l,m)}_{\Delta \mathbf{r}}(f) = \gamma_s \overline{B}(f) H^{(m)}_{\textrm{out}}(x_s,z_s,x_{\textrm{out}},z_{\textrm{out}},f)  H^{(l)}_{\textrm{in}}(x_{\textrm{in}},z_s,x_{\textrm{in}},z_{\textrm{in}},f). 
\end{equation}
where $H^{(l)}_{\textrm{in}}(x_s,z_s,x_{\textrm{in}},z_{\textrm{in}},f) $ and $H^{(m)}_{\textrm{out}}(x_s,z_s,x_{\textrm{out}},z_{\textrm{out}},f) $ stand for the $l^{\textrm{th}}$ and $m^{\textrm{th}}$ reverberation  order of the input and output PSFs, respectively. As before, each PSF can be, under the paraxial approximation, decomposed as a product of a frequency-invariant enveloppe and a depth-oscillating phase term  (see Appendix~\ref{B}):
\begin{equation}
\label{psf7}
H^{(n)}_{\textrm{in/out}}(\mathbf{r},\mathbf{r}_{\textrm{in/out}},f) =\underbrace{e^{i2\pi f \left ( \frac{z-L-z_{\textrm{in/out}}}{c} +\frac{(2n+1)L}{c_R} \right) }}_{\textrm{axial propagation}}\underbrace{H^{(0)}_{\textrm{out}} \left(x-x_{\textrm{in/out}},z-z^{\star (n)}_{\textrm{in/out}},f_c\right) }_{\textrm{transverse focusing}}
\end{equation}
with
\begin{equation}
\label{foc_depth_reverb}
    z^{(n)\star}_{\textrm{in/out}}=z_{\textrm{in/out}}+L-(2n+1)L\frac{c_R}{c},
\end{equation}
the depth of the focusing plane for the $l^{\textrm{th}}$ reverberation order. Injecting the Eq.~\ref{psf7} into Eq.~\ref{decompose} leads to the following expression for the coefficients of the time focused $\mathbf{R}-$matrix expressed in de-scan coordinates:
\begin{equation}
   {R}^{(l,m)}(\Delta x,\Delta z,x_{\textrm{in}},z_{\textrm{in}},\tau) \propto \underbrace{H_{\textrm{in}}^{(0)}\left(0,\Delta z_l^{\star} ,f_c\right)}_{\textrm{constant}} \underbrace{H_{\textrm{out}}^{(0)}\left(\Delta x,\Delta z^{\star}_m-\Delta z ,f_c\right)}_{\textrm{spatial focusing}} \underbrace{B\left( \tau+ \frac{\Delta z} {c} - \frac{2(l+m)L}{c_R}\right)}_{\textrm{time focusing}}.
   \label{Rbright3}
\end{equation}
with 
\begin{equation}
\label{refocus_depth}
  \Delta z^{\star}_{n} = 2nL \frac{c_R}{c} + L \left ( \frac{c_R}{c} - \frac{c}{c_R} \right ), 
\end{equation}
the de-scanned focusing depth of the $n^{\textrm{th}}$ reverberation order. 

The temporal term in Eq.~\ref{Rbright3} accounts for a time comb of multiply-reflected echoes behind the ballistic component, with a period equal to the round-trip time $2L/c_R$ inside the reverberating layer. The observed period of \textcolor{red}{2} $\mu$s in Fig.~\ref{fig5} is not consistent with reverberation in the Plexiglas layer that would imply a reverberation time of 4 $\mu$s.. It is more probably the signature of reverberation inside the acoustic lens ($c_r \sim 1000$ m/s and $L\sim 1$ mm) that lies between the probe and the Plexiglas. 

Each of the reverberated wave-front focuses at a de-scan depth $\Delta z^{\star}_m$ (Eq.~\ref{refocus_depth}) that only depends on the reverberation order $m$ of the output path. 
For the ballistic wave-front, the corresponding focusing time shift is therefore given by  
\begin{equation}
\label{refocus_time}\tau_0^{\star} = - \frac{\Delta z^{\star}_0}{c} = \frac{ L}{c_R} \left ( 1-\frac{c_R^2}{c^2}  \right )
\end{equation}
The measured value $\tau_0^{\star} \sim -5.5$ $\mu$s in Fig.~\ref{fig5} is this time consistent with a Plexiglas layer of thickness $L\sim 6$ mm. Indeed, the aberration undergone by the ballistic wave-front is mainly induced by the Plexiglas layer (and not by the acoustic lens that would yield $\tau_0^{\star} \sim -0.5$ $\mu$s). 

Through this example, we show that the reflection matrix is an ideal tool for analyzing the focusing process in the medium. By decoupling the imaging time from the ballistic echo time, it allows us to probe the temporal dispersion of the echoes and, for example, highlight the presence of multiple reflections. In the third article of the series~\cite{Giraudat2025b}, we will see that this information is valuable and can be used to compensate for multiple propagation paths. In that respect, another important feature highlighted by Eqs.~\ref{refocus_depth} and \ref{refocus_time} is the confirmation that the effect of the reverberating layer does not depend on the target depth $z_{\textrm{in}}$. All points below this layer experience the same focusing defect and reverberation phenomena. Such a an isoplanicity will be an important property to extract a self-portrait of the focusing process in the speckle regime.

Indeed, for the moment, we have limited ourselves to studying the focused reflection matrix associated with a bright scatterer. As pointed out in the introduction, real media do not contain isolated bright scatterers as in a phantom. For instance, in medical ultrasound, the biological tissues are mainly composed of a random distribution of sub-resolved scatterers generating ultrasonic speckle.  For this reason, in the next section, we will revisit the temporal study of the focused $\mathbf{R}$-matrix carried out in this section, extending it to the case of a speckle medium. 

\section{Self-portrait of a coherent wave packet in speckle}

In the speckle regime, the reflected wave-field is the result of random interference between echoes generated by un-resolved scatterers. An incident focal spot can thus be seen as an incoherent virtual source. While the wave-fields produced by such virtual sources are generally seen as spatially- and temporally-incoherent, we here show how to combine them in order to synthesize a coherent reflector and monitor the propagation of a coherent wave-packet despite disorder.

\subsection{Wave-packet associated with one virtual source}

\begin{figure*}[t!]
  \includegraphics[width=1\linewidth]{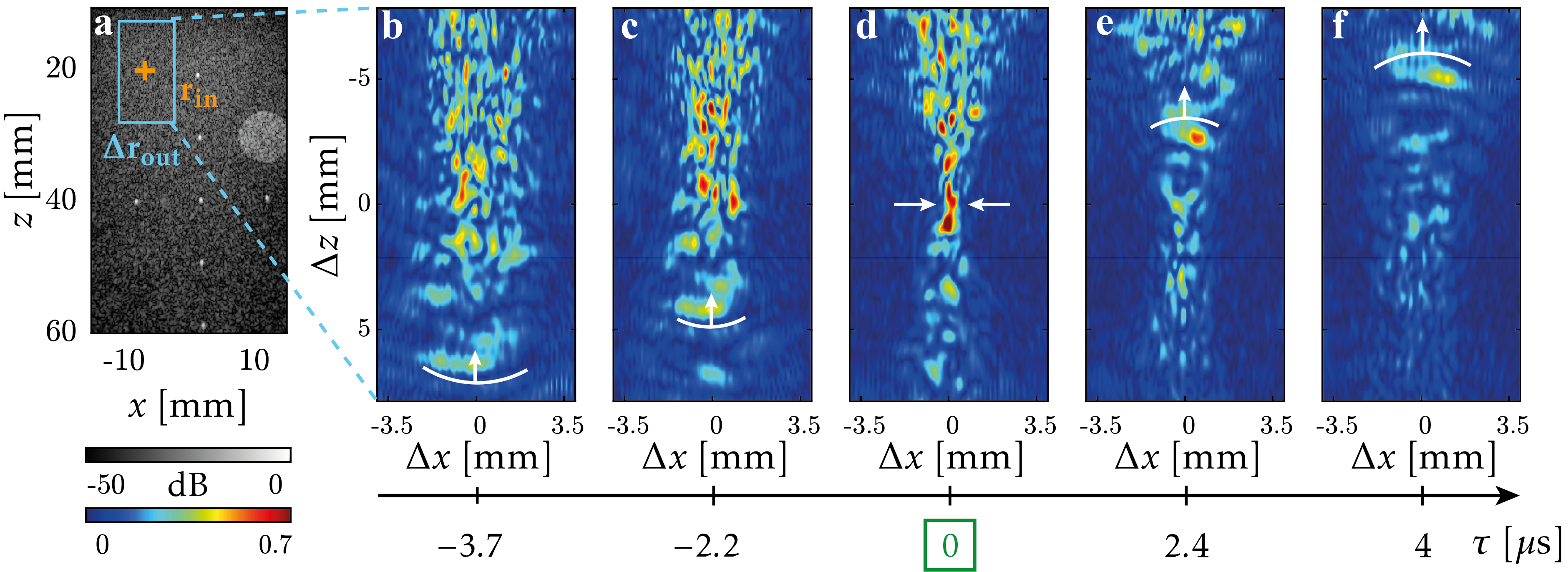}
    \caption{\textbf{Coherent wave-packet associated with a virtual source in the speckle region.}  \textbf{a}, Ultrasound image of the phantom (Fig.~\ref{fig1}e): The position of the virtual source $\mathbf{r}_{\textrm{in}}$ is indicated by an orange cross and the set of virtual receivers $\mathbf{r}_{\textrm{out}}$ surrounding it is delimited by a blue rectangle.
\textbf{b}-\textbf{f}, Normalized amplitude of the wave-field produced by the virtual source $\mathbf{r}_{\textrm{in}}$ in de-scanned coordinates displayed for various lapse times $\tau$. The propagation of the coherent wave-packet is also displayed in Supplementary Movie 3. The white arches highlight the coherent wave-front propagating upwards that can be discerned despite the random reflectivity of the speckle grains. }    
    \label{fig7}
\end{figure*}

We thus turn our attention to the more complex situation where the time-focused reflection matrix is constructed for a virtual source $\mathbf{r}_{\textrm{in}}$ and a set of virtual receivers $\mathbf{r}_{\textrm{out}}$ surrounding it, all located in a speckle zone, as shown in Fig.~\ref{fig7}a. First, we again consider the ideal case of the ultrasonic phantom experiment in which the velocity model used to form the matrix $\mathbf{R}_{\Delta \mathbf{r}}(\tau)$ is correct ($c=c_0$). Propagation of the wave induced by a given virtual source $\mathbf{r}_{\textrm{in}}$ in speckle can be displayed by reshaping the corresponding column of $\mathbf{R}_{\Delta \mathbf{r}}(\tau)$ in de-scanned coordinates $(\Delta x,\Delta z)$ at different times-of-flight $\tau$ (Fig.~\ref{fig7}b-f). 
As the reflected wave induced by a bright scatterer (Fig.~\ref{fig4}), a coherent wave propagating upwards seems to emerge (see the white arc in Fig.~\ref{fig7}b-f). The coherent beam seems to narrow to a maximum at the ballistic time ($\tau = 0$) and at the expected focusing point ($\Delta \mathbf{r} = \mathbf{0}$), suggesting its focus (white arrows in Fig.~\ref{fig7}d). However, the observation of the focusing process is hampered by the random reflectivity of the sub-resolved scatterers that modulates the amplitude of the back-propagated wave-field. Furthermore, the coherent wave is also partially masked by spurious echoes generated by out-of-plane scatterers located in the illumination cone, upstream and downstream of the virtual source, and, possibly, by multiply-scattered echoes. 

To circumvent this issue and isolate the coherent wave packet, one can take advantage of the spatial invariance of the problem and combine a set of propagation films associated with different virtual sources $\mathbf{r}_{\textrm{in}}$ belonging to the same isoplanatic patch.

\subsection{Extraction of the coherent wave packet by singular value decomposition}

\begin{figure*}[t!]
  \includegraphics[width=10cm]{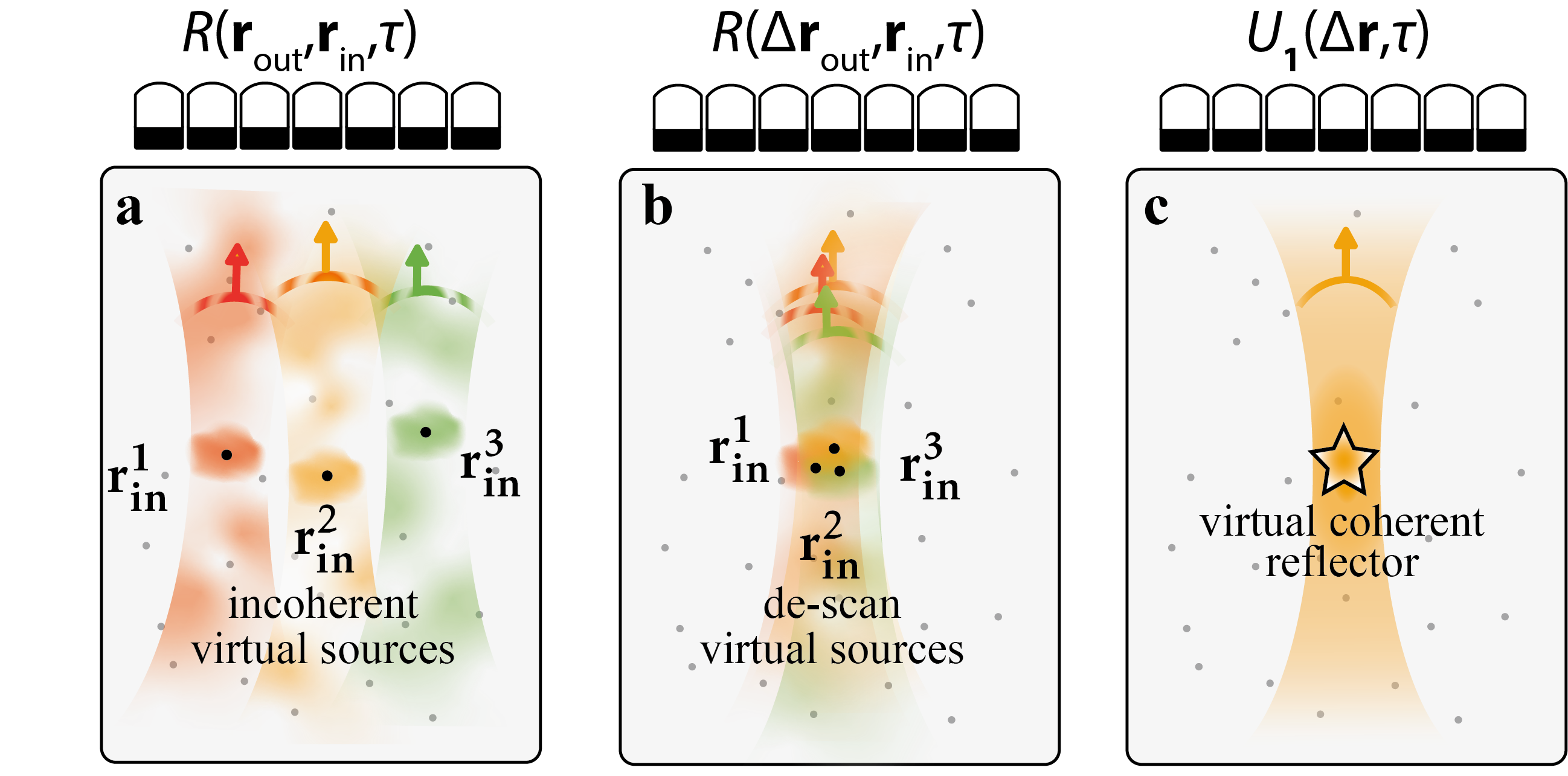}
    \caption{\textbf{Synthesis of a coherent guide star from a set of virtual sources $\mathbf{r}_{\textrm{in}}$ by iterative time reversal}. \textbf{a} The focused reflection matrix $\mathbf{R}_{\mathbf{rr}}(\tau)=[R(\mathbf{r}_{\textrm{out}},\mathbf{r}_{\textrm{in}},\tau)]$ contains the wave-fronts induced by a set of virtual sources synthesized in speckle.  \textbf{b} Express the reflection matrix in de-scan frame,  $\mathbf{R}_{\mathbf{\Delta r}}(\tau)=[R(\mathbf{r}_{\textrm{out}}-\mathbf{r}_{\textrm{in}},\mathbf{r}_{\textrm{in}},\tau)]$ , amounts to a lateral shift of virtual sources at the center of the field-of-view.  \textbf{c} An iterative time reversal process applied to $\mathbf{R}_{\mathbf{\Delta r}}(\tau)$ enables a rephasing of each virtual source in order to synthesize a coherent guide star. In practice, this operation can be performed by a SVD of the concatenated matrix $\mathbf{R}_{\lbrace \Delta \mathbf{r},\tau \rbrace \mathbf{r}}$. Its first eigenvector $\mathbf{U}_1=[U_1(\Delta \mathbf{r},\tau)]$ provides the coherent wave-field induced by this guide star.}    
    \label{fig8}
\end{figure*}

A set of virtual sources $\mathbf{r}_{\textrm{in}}$ belonging to the same isoplanatic patch are now considered (Fig.~\ref{fig8}a). Each of these sources should give rise to spatially correlated wave-fronts because of the shift-shift memory effect (Fig.~\ref{fig8}b). However, each virtual source is also modulated by the random distribution of sub-resolved scatterers lying into each incident focal spot. The idea is thus to exploit the spatial correlations between each focused wave-field in order to: (\textit{i}) synthesize a coherent guide star from these different realizations of disorder (Fig.~\ref{fig8}c); (\textit{ii}) extract a coherent wave packet equivalent to the one produced by a bright scatterer in Fig.~\ref{fig4}. Previous works have shown how a SVD process can be fruitful to synthesize such a coherent guide star~\cite{robert_greens_2008,lambert_distortion_2020,bendjador_svd_2020,lambert_ultrasound_2022}. However, those previous demonstrations were limited to a time-gated wave-field. Here, we show how this process can be extended to a time-dependent wave-field in order to extract a coherent spatio-temporal wave-packet in speckle. 

\begin{figure}[h!tb]\centering
   \includegraphics[width=\linewidth]{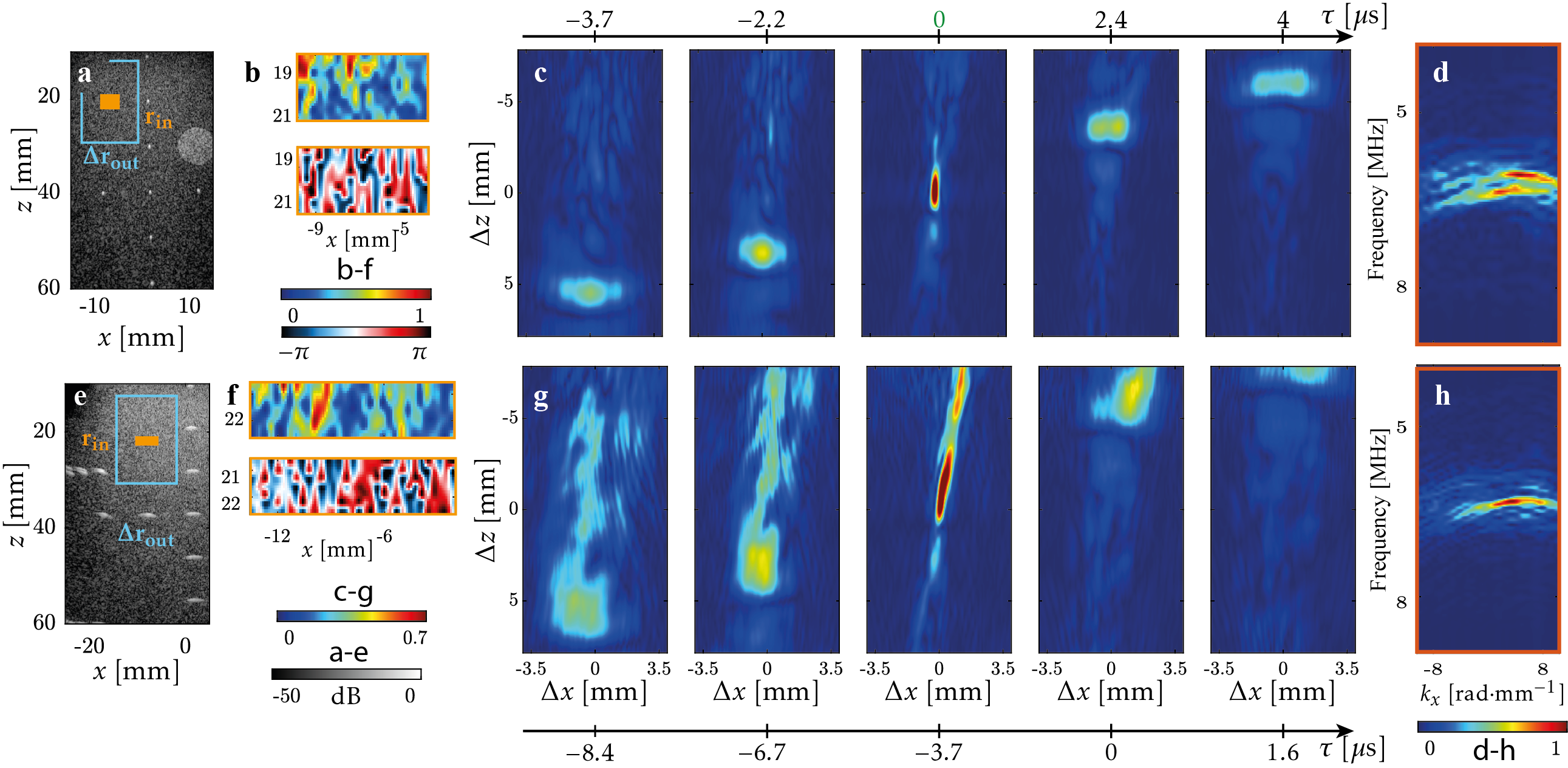}
   \caption{\textbf{Propagation movie associated with the coherent wave packed extracted by iterative time reversal from a set of virtual sources in the speckle.} \textbf{a}-\textbf{d} Correct speed-of-sound model ($c_0=1540$ m.s$^{-1}$): \textbf{a} The considered virtual sources $\mathbf{r}_{\textrm{in}} = (x_{\textrm{in}},z_{\textrm{in}})$ and their neighboring receivers are considered in the orange and blue rectangles on the ultrasound image.    \textbf{b} The input singular vector $\mathbf{V}_1$ yields the amplitude (top) and phase (bottom) of the weighting coefficient that shall be applied to each virtual source to synthesize the coherent guide star. \textbf{c} The first output singular vector $\mathbf{U}_1$ of the re-shaped 2D matrix $\mathbf{R}_{\lbrace \Delta \mathbf{r},\tau \rbrace \mathbf{r}}$ yields the propagation movie of the coherent wave-packet shown here in absolute value. \textbf{d}  Spatio-temporal Fourier transform of $\mathbf{U}_1$.  \textbf{e}-\textbf{h} Same as in panels a-d but for an incorrect wave velocity model ($c_0=1440$ m.s$^{-1}$)}
   \label{fig9}
\end{figure}

To do so, we first select the speckle area over which we want to evaluate the focusing process. This area should be as small as possible to guarantee a maximum isoplanicity. However, it should encompass a sufficiently large number of independent virtual sources in order to extract the coherent wave packet and get rid of speckle fluctuations, out-of-focus echoes and multiply-scattered waves. Selected areas of virtual sources are superimposed to the ultrasound images in Figs.~\ref{fig9}a and e for a correct and incorrect wave velocity model, respectively It should be noted that the isoplanicity condition for $c\neq c_0$ is more restrictive since here the velocity error induces aberrations that are not invariant with respect to depth $z$. For this reason, the selected area for an incorrect wave velocity model (Fig.~\ref{fig9}e) has been restricted axially and extended laterally compared to the reference case (Fig.~\ref{fig9}a). 

Once the area of virtual sources selected, the time-dependent de-scan reflection matrix is first reshaped in 2D by concatenating the dimensions $\Delta \mathbf{r} = \left( \Delta x, \Delta z\right)$ and $ \tau$: 
\begin{equation}
   \mathbf{R}_{\{\Delta \mathbf{r} ,\tau \},\mathbf{r}} = \left[ R(\{\Delta \mathbf{r}_{\textrm{out}}, \tau\},\mathbf{r}_{\textrm{in}})\right].
\end{equation}
Each column of the matrix corresponds to a spatio-temporal wave-field induced by a given virtual source $\mathbf{r}_{\textrm{in}}$ as the one displayed in Fig.~\ref{fig7}. To synthesize a coherent guide star from these different realizations over disorder, a singular value decomposition (SVD) of the concatenated matrix $ \mathbf{R}_{\{\Delta \mathbf{r} ,\tau \},\mathbf{r}}$ is performed such that: 
\begin{equation}
    \mathbf{R}_{\{\Delta \mathbf{r} ,\tau \},\mathbf{r}}  = \mathbf{U} \times \mathbf{\Sigma} \times \mathbf{V^\dag} 
\end{equation}
or, in terms of coefficients: 
\begin{equation}\label{ch3_R_SVDspeckle}
  R(\{\Delta \mathbf{r}_{\textrm{out}}, \tau\},\mathbf{r}_{\textrm{in}}) = \sum_p \sigma_p U_p(\Delta \mathbf{r}_{\textrm{out}},\tau)  V_p^*(\mathbf{r}_{\textrm{in}}). 
\end{equation}
The matrix $\Sigma$ is diagonal and its diagonal coefficients contains the singular values of $\mathbf{R}_{\{\Delta \mathbf{r} ,\tau \},\mathbf{r}} $ ranged in decreasing order ($\sigma_1> \cdots >\sigma_N$). The matrix $\mathbf{U}$ and $\mathbf{V}$ are unitary matrices whose columns $\mathbf{U}_p=[U_p(\Delta \mathbf{r}_{\textrm{out}},\tau) ]$ and $\mathbf{V}_p=[V_p(\mathbf{r}_{\textrm{in}})]$ are the singular vectors of $\mathbf{R}_{\{\Delta \mathbf{r} ,\tau \},\mathbf{r}} $.
Among all eigenstates, we are interested in the first one that maximizes the correlation of the wave-field induced by each virtual source. Figures~\ref{fig9}b-d show the first singular vectors of $\mathbf{R}_{\{\Delta \mathbf{r} ,\tau \},\mathbf{r}} $ for the area of virtual sources displayed in Fig.~\ref{fig9}a and beamformed using a correct wave velocity model ($c_0=c$). The first input eigenstate $\mathbf{V}_1$ (Fig.~\ref{fig9}b) contains the weighting coefficients that should be applied to each virtual source in order to synthesize a coherent guide star. The first output eigenstate $\mathbf{U}_1$ (Fig.~\ref{fig9}c) displays the wave-field generated by this guide star. Equivalent to the reflected wave-field induced by a bright scatterer (Fig.~\ref{fig4}c), this first result demonstrates our ability of synthesizing a guide star in speckle and probing the space-time evolution of the associated wave packet. The SVD has allowed us to overcome the random reflectivity of speckle grains and remove parasitic echoes observed in Fig.~\ref{fig7}b. A coherent wave propagating towards the probe is again observed. Nevertheless, the axial resolution of the wave packet seems to be degraded by the SVD process. This deterioration is even worse when a mismatch exists between the real speed-of-sound distribution in the medium and the wave velocity model (Fig.~\ref{fig9}g). Moreover, the focusing depth and time of the wave-packet seem to differ from the one induced by a bright scatterer in the latter case. To understand this discrepancy and the limits of the SVD process, an analytical study of this operation is now provided.

\subsection{Analytical expression of the focused reflection matrix in speckle}

The SVD result of Fig.~\ref{fig9} can be intuitively understood by considering the concatenated reflection matrix in the frequency domain, $\overline{\mathbf{R}}_{\lbrace \Delta\mathbf{r},f \rbrace \mathbf{r}}$. Indeed, the Fourier transform of Eq.~\ref{ch3_R_SVDspeckle} yields:
\begin{equation}\label{ch3_R_SVDspeckle2}
 \overline{R}(\{\Delta \mathbf{r}_{\textrm{out}}, f\},\mathbf{r}_{\textrm{in}}) = \sum_p \sigma_p \overline{U}_p(\Delta \mathbf{r}_{\textrm{out}},f)  V_p^*(\mathbf{r}_{\textrm{in}}). 
\end{equation}
Under a single scattering assumption, the coefficients of the monochromatic focused reflection matrices can be theoretically expressed as follows~\cite{lambert_reflection_2020,lambert_ultrasound_2022a}:  
\begin{equation}
\label{Rxx2}
   \overline{R}(x_{\textrm{out}},x_{\textrm{in}},z_{\textrm{out}},z_{\textrm{in}},f) = \iint d z d x~ H_{\textrm{out}}(x,z,x_{\textrm{out}},z_{\textrm{out}},f) \gamma(x,z) H_{\textrm{in}}(x,z,x_{\textrm{in}},z_{\textrm{in}},f),
\end {equation}
where $\gamma(x,z)$ represents the reflectivity of the medium at point $\mathbf{r} = (x,z)$. Injecting the paraxial expression of the input and output PSFs (Eq.~\ref{ch3_eq_monof_H}) into Eq.~\ref{Rxx2} leads to the following expression of the focused $\mathbf{R}-$matrix coefficients:
\begin{eqnarray}
   \overline{R}(x_{\textrm{out}},x_{\textrm{in}},z_{\textrm{out}},z_{\textrm{in}},f)&=& B(f) \iint dz dx~ H_{\textrm{out}}^{(0)}\left(x-x_{\textrm{out}},c(z-z^{\star}_{\textrm{out}})/c_0 ,f_c\right) \nonumber \\
  & \times &  \gamma(x,z) H_{\textrm{in}}^{(0)} \left(x-x_{\textrm{in}},c(z-z^{\star}_{\textrm{in}})/c_0,f_c\right)e^{i 2\pi f \left(\frac{ z_{\textrm{in}}+z_{\textrm{out}}}{c_0}-\frac{2z}{c} \right)}.
   \label{ch3_eq_monof_Rdrr}
\end{eqnarray}
In first approximation, the incident beam can also be described as perfectly focused at the focusing point $\mathbf{r}^{\star}_{\textrm{in}}=({x}_{\textrm{in}},z^{\star}_{\textrm{in}})$: 
\begin {equation}
\label{poinlikeguidestar}
   H_{\textrm{in}}^{(0)}\left(x-x_{\textrm{in}},c(z-z^{\star}_{\textrm{in}})/c_0,f_c\right)\sim   \delta(x-x_{\textrm{in}})\delta(z-z^{\star}_{\textrm{in}}),
\end{equation}
This approximation amounts to considering that the virtual source synthesized at $\mathbf{r}_{\textrm{in}} = (x_{\textrm{in}},z^{\star}_{\textrm{in}})$ as point-like. The expression of the focused reflection matrix in the concatenated de-scanned basis then becomes: 
\begin{equation}
   \overline{R}(\lbrace \Delta x,\Delta z,f \rbrace , \lbrace x_{\textrm{in}},z_{\textrm{in}} \rbrace ) \sim \underbrace{\gamma(x_{\textrm{in}},z_{\textrm{in}}^{\star})}_{V_1^*(x_{\textrm{in}},z_{\textrm{in}})} \underbrace{\overline{B}(f)H_{\textrm{out}}^{(0)}\left(-\Delta x,-\Delta z,f_c\right) e^{i 2\pi f \left [ \frac{ 2z_{\textrm{in}}}{c_0} \left (1-\frac{c_0^2}{c^2}\right)+\frac{\Delta z}{c_0} \right]}}_{\overline{U}_1\left(\Delta x,\Delta z,f\right)}.
\end{equation}
Comparison with Eq.~\ref{ch3_R_SVDspeckle2} shows that $\overline{\mathbf{R}}_{\lbrace \Delta \mathbf{r} , f \rbrace \mathbf{r}}$ is thus of rank 1.  On the one hand, the first input singular vector $\mathbf{V}_1$ directly provides the medium reflectivity at the focusing depth: $V_1(x_{\textrm{in}},z_{\textrm{in}})=\gamma(x_{\textrm{in}},z^{\star}_{\textrm{in}})$ (Figs.~\ref{fig9}b,f).  On the other hand, the first output singular vector directly provides the output PSF at the central frequency, $H_{\textrm{out}}^{(0)}\left(-\Delta x,-\Delta z,f_c\right)$, modulated by the frequency bandwidth of the recorded signals and a phase term that accounts for the axial propagation of the coherent wave-field. An inverse Fourier transform of the first eigenvector $\overline{\mathbf{U}}_1(f)$ provides an analytical expression for the coherent wave-packet extracted by the SVD process in the time domain: 
\begin{equation}
   U_1(\Delta x,\Delta z,\tau) = H_{\textrm{out}}^{0}\left(-\Delta x,-\Delta z,f_c \right) B\left( \tau + 2\tau^{\star} +\frac{\Delta z}{c_0} \right).
   \end{equation}
This result confirms that the vector $\mathbf{U_1}$ provides access to a coherent spatio-temporal wave-packet in a speckle environment with a spatial envelope accounting for the output PSF at the central frequency and a time shift $2\tau^{\star}$ (Eq.~\ref{tau_star}) accounting for the time delay experienced by the incident and reflected waves with respect to the ballistic time expected under the wave velocity assumption.

These analytical predictions are in good agreement with the spatio-temporal wave packets displayed in Figs.~\ref{fig9}c and g. For a correct wave velocity model ($c_0=1540$ m.s$^{-1}$, Fig.~\ref{fig9}c) , the wave-packet actually focuses at the origin ($\Delta z=0$) and at the expected ballistic time  ($\tau =0$). For an incorrect wave velocity model ($c_0=1440$ m.s$^{-1}$, Fig.~\ref{fig9}g), the focus is still observed at $\Delta z=0$ mm and close to the expected focusing time ($2\tau^{\star} \sim 3.7$ $\mu$s). Nevertheless, the point-like virtual source hypothesis made for analytical tractability does not provide a clue for the loss of axial resolution observed in Figs.~\ref{fig9}c and g. 

To go beyond this analytical calculation, one can exploit the analogy between the SVD and iterative time reversal processing~\cite{lambert_distortion_2020,lambert_ultrasound_2022}. The first eigenstate can be seen as the result of an iterative time reversal experiment that tends to maximize the energy back-scattered by our virtual guide star. As the virtual source is, in reality, not point-like but displays a depth extension dictated by the depth-of-field $\delta z \sim 7 \lambda (z_\textrm{in}/D)^2$ (with $D$ the array extension), the SVD process converges towards a wave-front of limited frequency-bandwdith $\delta f \sim c /(2 \delta z)  $ around the central frequency $f_c$. The first time-reversal invariant extracted by the SVD process optimizes the energy coming from the elongated guide star. This feature is confirmed by the spatio-temporal Fourier transform of the first eigenvector $\mathbf{U}_1$. While the angular distribution of the wave-field is quite homogeneous in the transverse spatial frequency domain, its energy is peaked around the central temporal frequency $f_c$ (Figs.~\ref{fig9}d,h). The bandwidth $\delta f$ is even more limited when the wave velocity model is not correct (Figs.~\ref{fig9}h). Indeed, the axial distortions undergone  by the incident focal spot leads to an axial extension of the guide star which, in turn, reduces the bandwidth of the first output singular vector $\mathbf{U}_1$.

\subsection{Extraction of the coherent wave packet by iterative phase reversal}

\begin{figure}[h!tb]\centering
   \includegraphics[width=\linewidth]{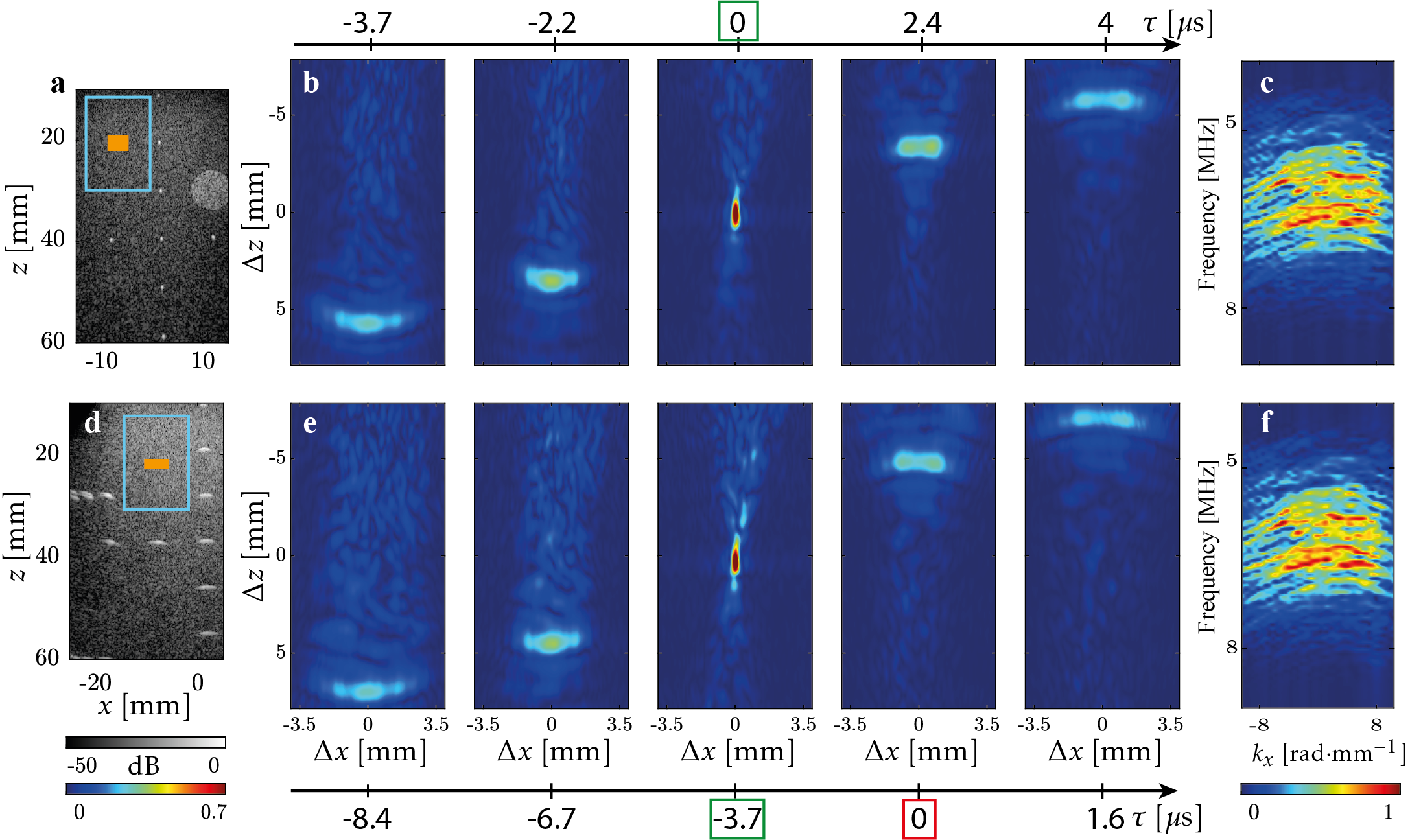}
   \caption{\textbf{Propagation movie associated with the coherent wave packet extracted by iterative phase reversal from a set of virtual sources in the speckle.} Correct wave velocity model ($c_0=1540$ m.s$^{-1}$): \textbf{a} The considered virtual sources $\mathbf{r}_{\textrm{in}} = (x_{\textrm{in}},z_{\textrm{in}})$ and their neighboring receivers are considered in the orange and blue rectangles on the ultrasound image. \textbf{b} The iterative phase
reversal invariant yields the propagation movie of the coherent wave-packet $T(\Delta x,\Delta z,\tau)$  shown
here in absolute value at different lapse times $\tau$. \textbf{c} Frequency-angular spectrum $\overline{\overline{T}}(k_x,\Delta z,\tau)$ of the iterative phase reversal invariant. \textbf{d}-\textbf{f} Same as in panels a-c but for an incorrect wave velocity model ($c_0=1440$ m.s$^{-1}$).}
   \label{fig10}
\end{figure}

To circumvent that issue, an iterative phase reversal process can be operated from the $\mathbf{k}-$space thanks to the distortion matrix concept~\cite{lambert_distortion_2020}. Originally proposed for a time-gated reflection matrix~\cite{bureau_three-dimensional_2023}, it is here applied to the broadband reflection matrix. To this end, the reflection matrix is projected from the de-scanned basis to the plane wave basis $(k_x)$ at each frequency $f$. This operation gives direct access to the distortion matrix in the spatio-temporal Fourier domain:
\begin {equation}
   D\left(k_{x},\Delta z,\mathbf{r}_{\textrm{in}},f\right) = \iint  dt d \Delta x  ~ R\left(\Delta x,\Delta z,\mathbf{r}_{\textrm{in}},\tau \right) e^{i k_{x}.\Delta x}  e^{-i 2 \pi f \tau}.
\end{equation}
The distortion matrix contains the deviation of each angular component of reflected waves from an ideal wavefront that would be obtained in the absence of aberrations and without multiple scattering. To synthesize a coherent guide star from the set of virtual sources $\mathbf{r}_{\textrm{in}}$, each distorted wave-front shall be first correlated with each other. Mathematically, this operation consists in computing a correlation matrix $\mathbf{C}=[ C\left( \{k_{x},\Delta z,f\},\{k^\prime_{x},\Delta z^\prime,f^\prime\} \right)]$ between each angular/depth/frequency $(k_{x}/\Delta z/f)$ component of the distorted wave-fronts:
\begin{equation}
   C\left( \{k_{x},\Delta z,f\},\{k^\prime_{x},\Delta z^\prime,f^\prime\} \right) = \sum_{\mathbf{r}_{\textrm{in}} } D(k_{x},\Delta z,f,\mathbf{r}_{\textrm{in}}) D^*(k_{x}^{\prime},\Delta z_{\textrm{out}}^\prime,f^\prime,\mathbf{r}^\prime_{\textrm{in}}).\label {ch3_eq_Cout}
\end{equation}
The first singular vector $\overline{\overline{\mathbf{U}}}_1$ of the distortion matrix, or equivalently, the first eigenvector of $\mathbf{C}$, is the invariant of a time reversal process applied to $\mathbf{D}$, that can be expressed as follows:
\begin{equation}
\label{ITR}
    \sigma_1^{(n)}\overline{\overline{\mathbf{U}}}^{(n)}_1=\mathbf{C}\times \overline{\overline{\mathbf{U}}}_1^{(n-1)}.
\end{equation}
with $\overline{\overline{\mathbf{U}}}_1 =  \lim\limits_{\mathrm{n} \to \infty} \overline{\overline{\mathbf{U}}}_1^{(n)}$ and $\sigma_1 =  \lim\limits_{\mathrm{n} \to \infty} \sigma_1^{(n)}$. As we have seen before, such an iterative time reversal process is not optimal since it tends to maximize the back-scattered energy and therefore concentrates on the most energetic part of the frequency bandwidth (Figs.~\ref{fig9}d,h). To circumvent this issue, Eq.~\ref{ITR} can be slightly modified to converge towards a wave-front effective over the whole frequency bandwidth and angular spectrum, such that
\begin{equation}
\label{IPR}
 \mathbf{W}^{(n+1)} = \exp \left( i \arg \left \{  \mathbf{C} \times \mathbf{W}^{(n)} \right  \} \right),
\end{equation}
with $\mathbf{W}^{(0)}  = \left[1 \cdots 1\right]^\top$, an arbitrarily chosen unit vector. The phase reversal invariant, $\mathbf{W} =  \lim\limits_{n\to \infty} \mathbf{W}^{(n)}$, can be applied a last time to the correlation matrix in order to recover the frequency and angular spectrum of the wave-field:
\begin{equation}
  \overline{\overline{\mathbf{T}}} =  \mathbf{C} \times \mathbf{W}
\end{equation}
The resulting wave-field $\overline{\overline{\mathbf{T}}}$ is an estimator of the medium transmittance in the space-time Fourier space. In contrast with the SVD process (Figs.~\ref{fig9}d and h), the IPR algorithm converges towards an invariant that now spans over the whole frequency bandwidth, here 5-8 MHz (Figs.~\ref{fig10}c and f).

Inverse Fourier transforms in the $(f,k)-$space finally yield the movie of the coherent wave-packet in the spatial and time domains:
\begin{equation}
   {T}\left(\Delta x,\Delta z,\tau\right) =   \iint df dk_x ~  \overline{\overline{T}}\left(k_x,\Delta z,f\right) e^{-i k_x.\Delta x}  e^{i 2 \pi f \tau}.
   \label{PSFt}
\end{equation}
The result is displayed in Fig.~\ref{fig10}b for a correct wave velocity model $c=c_0$. Comparison with the coherent wave packet resulting from the SVD (Fig.~\ref{fig9}c) shows the drastic gain in axial resolution provided by the IPR process. Its benefit is even more striking in the aberrated case. While the SVD wave packet spreads over a few millimeters and exhibits a low contrast with respect to spurious echoes (Fig.~\ref{fig9}g), the IPR wave-packet is highly resolved in time and space (Fig.~\ref{fig10}e). This feature is explained by the much broader frequency bandwidth of the coherent wave packet with IPR (Fig.~\ref{fig10}f) compared to SVD (Fig.~\ref{fig9}h).  The estimation of the focusing time can now be made with high precision and could lead to a sharp estimation of the speed-of-sound: $\hat{c}\sim 1545$ m.s$^{-1}$ for $2\tau^{\star} \simeq - 3.7$ $\mu$s (Eq.~\ref{tau_star}).

Nevertheless, we will see, in the second article of the series~\cite{Bureau2024}, how we can adopt a more direct and robust strategy for local speed-of-sound estimation. Indeed, one of the limit of our method here is that it requires a region of homogeneous speckle. Moreover, this area should be large enough to encompass a sufficient number of independent realizations of disorder. The bias made on the transmittance estimator actually scales as: 
\begin{equation}
\label{bias}
   || \delta \overline{\overline{T}}  ||^2 \propto \frac{1}{\mathcal{C}^2 N_{\textrm{in}}},
\end{equation}
where $N_{{\textrm{in}}}$ is the number of diffraction-limited resolution cells in the selected area of virtual sources. $\mathcal{C}$ is a coherence factor that is an indicator of the quality of focusing \cite{mallart_adaptive_1994}, defined as the ratio between the coherent intensity and the average incoherent intensity of the back-scattered echoes:
\begin{equation}
    \mathcal{C}=\frac{\left | \sum_{k_x}\sum_f D(k_x,\Delta z,f) \right |^2}{ \sum_{k_x}\sum_f \left |D(k_x,\Delta z,f) \right |^2}
\end{equation}
Physically, the bias error $|| \delta \overline{\overline{T}} ||^2 $ inversely proportional to $\mathcal{C}^2$ (Eq.~\ref{bias}) reflect the following effect: the more degraded the focusing and the more blurred the guide star, the greater the bias on the estimator. One can actually show that the coherence factor is inversely proportional with respect to the number of independent spatio-temporal degrees-of-freedom exhibited by the distorted wave-front. Their order of magnitude scales as the inverse of the number of resolution cells exhibited by the focal spot along each of its dimension:
\begin{equation}
    \mathcal{C}\sim \frac{\delta x_0}{\delta x}\frac{\delta z_0}{\delta z}\frac{\delta t_0}{\delta t}.
\end{equation}
The ratio $\delta x/ \delta x_0$ corresponds to the number of resolution cells exhibited by the focal spot along the lateral dimension, \textit{i.e} the ratio between the lateral extension  $\delta x$ of the focal spot and the diffraction-limited resolution length $\delta x_0 \sim \lambda/2 \sin \theta$, with $\theta$ the angle under which each virtual source sees the transducers array. The ratio $\delta z/ \delta z_0$ corresponds to the number of resolution cells exhibited by the focal spot along the axial dimension, \textit{i.e} the ratio between the axial extension $\delta z$ of the focal spot and $\delta z_0 \sim 2 \lambda/ \sin^2 \theta$, the depth-of-field at the position of virtual sources. At last, the ratio $\delta t/\delta_0$ corresponds to the number of time samplings exhibited by the focal spot in the time domain, with $\delta t$ the time spreading of the focal spot and $\delta t_0 \sim \Delta f^{-1}$, the time resolution of our measurement and $\Delta f$, the frequency bandwidth.  

On the one hand, Equation~\ref{bias} indicates that the number of independent virtual sources should be quite large to obtain a satisfying estimation of the wave-field, especially when the velocity model is far from reality (low coherence factor). On the other hand, all these virtual sources should belong to the same isoplanatic patch, the size of the latter being very reduced in presence of distributed aberrations.  This dilemma can be circumvented by reducing the space of solutions. One can consider, for instance, the time-gated reflection matrix and only focus on the spatial degrees of freedom of the wave-field. This strategy can be followed in absence of multiple reflections, as we will show in the second article of the series~\cite{Bureau2024} for compensating the axial distortions of a confocal image due to speed-of-sound variations.

\subsection{Reducing the space of solutions in presence of temporal dispersion}

In presence of temporal dispersion, an alternative strategy is to first focus on the lateral/frequency degrees of freedom before retrieving, in a second step, the axial evolution of the wave-field. Indeed, the dispersion relation implies a connection between the axial and traverse component of the wave vector and the frequency: $k_x^2+k_z^2=(2 \pi f/c)^2$. Reducing the space of solutions to the plane $(k_x,f)$ is therefore not detrimental and can facilitate the convergence of the IPR process.  
\begin{figure}[h!tb]\centering
   \includegraphics[width=\linewidth]{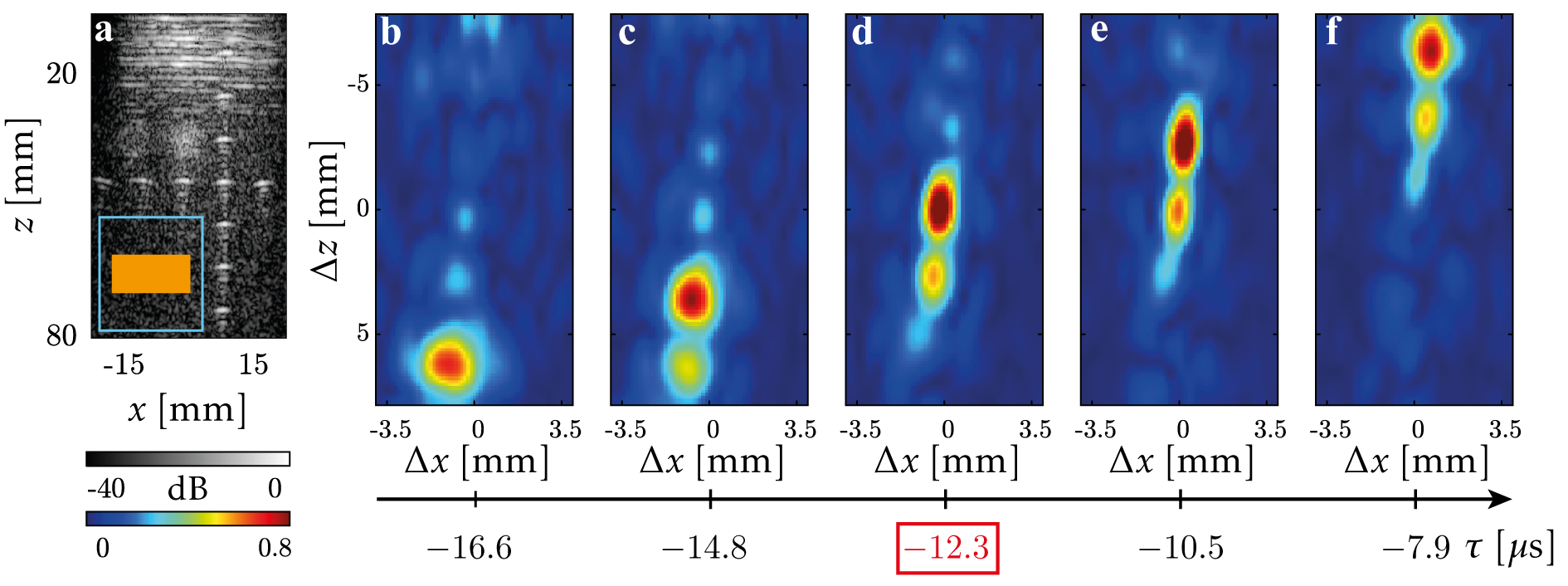}
   \caption{\textbf{Propagation movie associated with the coherent wave packet extracted by iterative phase reversal from a set of virtual sources in speckle in presence of reverberations.} \textbf{a} The considered virtual sources $\mathbf{r}_{\textrm{in}} = (x_{\textrm{in}},z_{\textrm{in}})$ and their neighboring receivers are considered in the orange and blue rectangles on the ultrasound image. \textbf{b}-\textbf{f} The iterative phase reversal wave-front $T(\Delta x ,\Delta z ,\tau)$ yields the propagation movie of the coherent wave-packet shown here in absolute value at different lapse times $\tau$.}
   \label{fig11}
\end{figure}

This is actually the case for the reverberating plate experiment previously described in Fig.~\ref{fig1}e. The IPR process described in Eq.~\ref{IPR} is unable to converge when applied to the focused reflection matrix for virtual sources contained in the speckle area displayed in Fig.\ref{fig11}a. To circumvent this problem, we have restricted the IPR analysis to the cross-section of the distortion matrix at $\Delta z=0$:  $\mathbf{D}_{\{k,f\},\mathbf{r}}(\Delta z=0)=[D(\{k_x,f\},\Delta z=0,\mathbf{r}_{\textrm{in}})]$. The result is a reverberation transmittance, $\overline{\overline{\mathbf{T}}}(\Delta z=0)=[\overline{\overline{T}}(k_x,f,\Delta z=0)]$, restricted to a de-scan depth $\Delta z=0$ and displayed in Fig.~5d of Ref.~\cite{Giraudat2025b}. 

To retrieve the axial propagation of this wave-front along the $z-$axis, a hint is to extract the weighting vector that shall be applied to the set of virtual sources $\mathbf{r}_{\textrm{in}}$ to synthesize the coherent guide star. This vector can be retrieved by considering the product between the  reduced distortion matrix and the corresponding transmittance such that:
\begin{equation}
    \mathbf{V}=\overline{\overline{\mathbf{T}}}^{\dag}(\Delta z=0) \times \mathbf{D}_{\{k,f\},\mathbf{r}}(\Delta z=0)
\end{equation}
The phase conjugate of the resulting vector $\mathbf{V}=[V(\mathbf{r}_{\textrm{in}})]$ contains the weighting coefficients that shall be applied to the matrix $\mathbf{D}_{\{k,f\},\mathbf{r}}$ in order synthesize the coherent guide star and retrieve an aberration transmittance for each de-scan depth:
\begin{equation}
   \overline{\overline{\mathbf{T}}}(\Delta z)=  \mathbf{D}_{\{k,f\},\mathbf{r}}(\Delta z) \times \mathbf{V}^{\dag}. 
\end{equation}
The spatio-temporal evolution of the corresponding wave-front can be deduced by a two-dimensional inverse Fourier transform (Eq.~\ref{PSFt}). The result is displayed at different time lapses in Fig.\ref{fig11}b. As the wave-front induced by a bright scatterer (Fig.~\ref{fig5}), the synthetic guide star reveals a tail of multiple reflections generated by the acoustic lens and exhibiting a time period of 2 $\mu$s. The time arrivals and focusing depths of the detected echoes, however, differ with the wave-fronts generated by a bright scatterer. 

\subsection{Analytical expression of the coherent wave-front in presence of reverberations}

To explain this difference, a theoretical expression of the estimated wave-front can be derived. In speckle and under a single scattering assumption, each reverberated component of the focused reflection matrix can be actually be expressed as follows:  
\begin{equation}
\label{Rxx3}
   \overline{R}^{(l,m)}(x_{\textrm{out}},x_{\textrm{in}},z_{\textrm{out}},z_{\textrm{in}},f) = \iint d z d x~ H^{(m)}_{\textrm{out}}(x,z,x_{\textrm{out}},z_{\textrm{out}},f) \gamma(x,z) H^{(l)}_{\textrm{in}}(x,z,x_{\textrm{in}},z_{\textrm{in}},f).
\end {equation}
Injecting the expression of the reverberated PSF  (Eq.~\ref{psf7}) leads to the following expression:

 \begin{eqnarray}
     \overline{R}^{(l,m)}(x_{\textrm{out}},x_{\textrm{in}},z_{\textrm{out}},z_{\textrm{in}},f) &=& B(f)\iint dz dx~ H_{\textrm{out}}^{(0)}\left(x-x_{\textrm{out}},z-z^{(m)\star}_{\textrm{out}})/c_0 ,f_c\right) \gamma(x,z)  \nonumber \\
  & \times &  H_{\textrm{in}}^{(0)} \left(x-x_{\textrm{in}},z-z^{(l)\star}_{\textrm{in}},f_c\right)e^{i2\pi f \left ( \frac{2(z-L-z_{\textrm{in}})-\Delta z}{c} +\frac{2(l+m+1)L}{c_R} \right) }.
 \end{eqnarray}
 As before, the virtual source is synthesized at the ballistic focusing point $\mathbf{r}^{\star}_{\textrm{in}}=({x}_{\textrm{in}},z^{(0)\star}_{\textrm{in}})$, with $  z^{(0)\star}_{\textrm{in/out}}=z_{\textrm{in/out}}+L\left (1-\frac{c_R}{c} \right)$ (Eq.~\ref{foc_depth_reverb}). In first approximation, the guide star position can be taken into account by assuming a point-like input focused beam at this point (Eq.~\ref{poinlikeguidestar}): 
\begin{eqnarray}
   \overline{R}^{(l,m)}(\lbrace \Delta x,\Delta z,f \rbrace , \lbrace x_{\textrm{in}},z_{\textrm{in}} \rbrace ) &\sim & \underbrace{\gamma(x_{\textrm{in}},z_{\textrm{in}}^{(0)\star})}_{V^*(x_{\textrm{in}},z_{\textrm{in}})} \nonumber \\ &\times & \underbrace{\overline{B}(f)H_{\textrm{out}}^{(0)}\left(-\Delta x,\Delta z^{\star}_m+\Delta z,f_c\right) e^{i2\pi f\left [ 2\frac{L}{c}\left (\frac{c}{c_R}- \frac{c_R}{c} \right )+ \frac{2(l+m)L}{c_R}-\frac{\Delta z}{c} \right ]}}_{T^{(l,m)}\left(\Delta x,\Delta z,f\right)}.
   \label{transmittance}
\end{eqnarray}
with 
\begin{equation}
\label{new_refocus_depth}
\Delta z_m^{\star}=- 2mL\frac{c_R}{c},
\end{equation}
the de-scan focusing depth of the $m^{\textrm{th}}$ reverberation order. Each component of the reverberated wave-front focuses at a de-scan depth $\Delta z^{\star}_m$ (Eq.~\ref{new_refocus_depth}) that only depends on the reverberation order $m$ of the output path. The comparison betwwen Eqs.~\ref{refocus_depth} and \ref{new_refocus_depth} show the difference of focusing de-scan depths for an isolated scatterer and in speckle. Contrary to the wave-field produced by a bright target whose ballistic echo shows a defocus, we retrieve the fact that, in the speckle regime, the ballistic wave-front should focus at $\Delta z=0$ (Fig.~\ref{fig11}d). This is a manifestation of a virtual guide star synthesized, not at the expected depth $z_{\textrm{in}}$, but in the focusing plane $z^{\star}_{\textrm{in}}$. Interestingly, each multiply-reflected component shows a defocus proportional to the number of reflections undergone by the output path, an effect also visible for the second and third order multiple reflection events in Figs.~\ref{fig11}e and f, respectively. 

An inverse Fourier transform of the reverberation transmittance $T(\Delta x,\Delta z,f)$ in Eq.~\ref{transmittance} provides an analytical expression for each multiply-reflected component of the coherent wave-packet extracted by the IPR process in the time domain: 
\begin{equation}
\label{wave_front_speckle}
   T^{(l,m)}(\Delta x,\Delta z,\tau) = H_{\textrm{out}}^{0}\left(-\Delta x,\Delta z^{\star}_m-\Delta z,f_c\right) B\left( \tau +\frac{\Delta z}{c}-  \tau_{l,m}^{\star} \right),
   \end{equation}
   with 
   \begin{equation}
    \label{time_shift}
        \tau_{l,m}^{\star}=2\tau_0^{\star}+ \frac{2(l+m)L}{c_R},
   \end{equation}
the time shift exhibited by each multiply-reflected echo. The first term of the latter equation corresponds to the time delay experienced by the ballistic wave while the second term is the time delay accumulated by each multiply-reflected wave inside it. Equation~\ref{wave_front_speckle} shows that, in a speckle environment, the IPR process provides a coherent wave-packet whose spatial envelope accounts for the output PSF at the central frequency and a time shift $\tau_{l,m}^{\star}$ that accounts for the delay experienced by each reverberated wave-front on the way-and-return path. Equation~\ref{time_shift} confirms theoretically that the ballistic time delay observed with a virtual guide star in speckle ($ \tau_{0,0}^{\star}\sim -12 $ $\mu$s in Fig.~\ref{fig11}d) is roughly two times the time delay obtained for a bright scatterer ($ \tau_{0}^{\star}\sim -5.5 $ $\mu$s Fig.~\ref{fig5}d). This difference is explained by the fact that, while the bright scatterer remains defocus in the latter case, the virtual guide star is synthesized in the real focal plane in the former case. This is an important feature since the size of the guide star has also an impact on the quality of the estimated transmittance. 

The speckle regime thus offers an ideal framework for aberration and reverberation compensation. In the third paper~\cite{Giraudat2025b} of the series, we will show how time matrix imaging can be exploited to tailor complex spatio-temporal focusing laws that enable a fine monitoring of the interference between the ballistic wave and the multiply-reflected paths.

\section{Discussion}

In this paper, we introduced the time-focused reflection matrix. Through the various examples discussed, this matrix is shown to provide an access at each point in the medium to a spatio-temporal self-portrait of the focusing process. This allows us to locally probe focusing defects due to wave velocity variations and the temporal spreading of echoes caused by multiple reflections, frequency dispersion, or multiple scattering. Besides direct focusing and imaging applications, the self-portrait of the coherent wave also contains a wealth of information to perform a better characterization of the medium. The arrival time and focusing depth of the echoes can be exploited for determining the thickness and speed-of-sound of the different layers on top of the imaging volume. At a longer term, this observable could also be included in the cost function of an optimization scheme that would consist in retrieving the speed-of-sound distribution inside the medium by minimizing the axial distortions exhibited by the coherent wave packet. In presence of reverberations, this would, however, require a forward model accounting for multiple scattering phenomena.

In that respect, a more direct route to speed-of-sound tomography will be investigated in the second paper~\cite{Bureau2024} of the series. Instead of probing the axial distortions of the focused wave-field in the time domain, the idea will be to investigate the self-portrait at the expected ballistic time as a function of the speed-of-sound model. Such an approach will allow us to circumvent the requirement of an homogeneous region of speckle to get the self-portrait of the wave-field. This condition can actually be quite restricting in view of applications. Moreover, we will still be able to probe coherent phenomena in speckle such as the Gouy phase in order to compensate the axial distortions of images induced by wave velocity fluctuations.

One remaining question is the possibility of performing a self-portrait of the wave-field inside more complex situations than the academic experiments of the current paper. Here, we considered experimental configurations with a lateral invariance that guaranteed a long range shift-shift memory effect~\cite{osnabrugge_generalized_2017} and our ability to synthesize a virtual guide star from a set of incoherent virtual sources. In more realistic situations, this isoplanatic range can become very reduced and the convergence of the IPR process operated form the k-space would no longer be guaranteed. In such configurations, a solution would be to operate the IPR process from a correction basis adapted to the shape and location of the reverberating layers inside the medium. As shown by adaptive focusing, a correction plane conjugated with the aberrating layer actually maximizes the isoplanicity of the correction.

To conclude, this work demonstrates the ability of a spatio-temporal control of the wave-field inside complex media. The originality of this study lies in the fact that this control is done from the only knowledge of the reflection matrix, without any detector inside the medium. Using the random distribution of scatterers as virtual microphones, we are able to probe the spatio-temporal evolution of the wave inside the medium. As we will show in follow-up studies~\cite{Bureau2024,Giraudat2025b}, this observable can be leveraged for a spatio-temporal compensation of aberrations and reverberations that usually induce strong distortions. Although the experimental proof-of-concept has been provided with ultrasound, the self-portrait concept can be potentially extended to all fields of wave physics in which a reflection matrix can be measured. The cellular components such as lipids or mitochondria at a micron-scale~\cite{Barolle2024} or the complex network of stacked magma lenses lying below a volcano~\cite{giraudat_unveiling_2023} generate a speckle wave-field that can be exploited for producing a self-portrait of the focusing process for light or seismic waves. 

\newpage

\noindent\textbf{Acknowledgments.}
The authors are grateful for the funding provided by the European Research Council (ERC) under the European Union's Horizon 2020 research and innovation program (grant agreement 819261, REMINISCENCE project, AA). W.L. acknowledges financial support from the SuperSonic Imagine company.\\

\noindent\textbf{Author contributions.}
A.A. and M.F. initiated the project. A.A. supervised the project. F.B. and W.L. performed the experiments. E.G., F.B. and W.L. developed the post-processing tools for the time-focused reflection matrix. E.G. and A.A. performed the theoretical analysis.  E.G. and A.A. prepared the manuscript. E.G., F.B., W.L., M.F. and A.A. discussed the results and contributed to finalizing the manuscript.

\clearpage 

\appendix

\section{Theoretical expression for the point spread function in free space}
\label{A}
In this Appendix, we want to express analytically the focused wave-field $H(\mathbf{r}_{s},\mathbf{r}_{\textrm{in}},f)$ at point $\mathbf{r}_{s}$  frequency $f$ when trying to focus at a point $\mathbf{r}_{\textrm{in}}$ assuming an homogeneous speed-of-sound $c_0$ inside the medium. By virtue of Rayleigh Sommerfeld integral, this wave field can be expressed as follows:
\begin{equation}
\label{psf}
H(\mathbf{r},\mathbf{r}_{\textrm{in}},f) = \sum_u G^*_0(\mathbf{r}_{\textrm{in}}-\mathbf{u},f)  \partial_z G(\mathbf{r}-\mathbf{u},f) 
\end{equation}
where $G(\mathbf{u},\mathbf{r}_{\textrm{in}},f) $ is the ballistic Green's function inside the medium that displays a background speed-of-sound $c$.  $G^*_0(\mathbf{u},\mathbf{r}_{\textrm{in}},f) $ corresponds to the incident wave-field expressed at the probe. It corresponds to the phase conjugate of the Green's function between each transducer $\mathbf{u}$ and the targeted focusing point $\mathbf{r}_{\textrm{in}}$. To go further, the 1D Fourier decomposition of the Green's function can be used:
\begin{equation}
    G(x,z,f)=\int  d k_x \frac{1}{2j}\frac{1}{\sqrt{\left (\frac{2\pi f}{c} \right )^2-k_x^2}} \exp \left (i z \sqrt{\left (\frac{2\pi f}{c} \right )^2-k_x^2}\right ) \exp(i k_x x)
\end{equation}
Injecting the latter equation into Eq.~\ref{psf} leads to:
\begin{equation}
\label{psf2}
H(\mathbf{r},\mathbf{r}_{\textrm{in}},f) = \frac{1}{2} \sum_u \int  d k_{\textrm{in}} \int d k_x \frac{e^{i z  \sqrt{ \left (\frac{2\pi f}{c} \right )^2-k_x^2} }e^{-i z_{\textrm{in}}  \sqrt{ \left (\frac{2\pi f}{c_0} \right )^2-k_x^2} }}{\sqrt{\left (\frac{2\pi f}{c_0} \right )^2-k_{\textrm{in}}^2}}   e^{i k_x (x-u)} e^{-i k_{\textrm{in}} (x_{\textrm{in}}-u)}
\end{equation}
The sum over transducers yields, in first approximation, the Dirac distribution $\delta(k_x -k_{\textrm{in}} )$. Equation~\ref{psf2} simplifies into:
\begin{equation}
\label{psf3}
H(\mathbf{r},\mathbf{r}_{\textrm{in}},f) = \frac{1}{2} \int  d k_{\textrm{in}} \frac{e^{i z  \sqrt{ \left (\frac{2\pi f}{c} \right )^2-k_{\textrm{in}}^2} }e^{-i z_{\textrm{in}}  \sqrt{ \left (\frac{2\pi f}{c_0} \right )^2-k_{\textrm{in}}^2} }}{\sqrt{\left (\frac{2\pi f}{c_0} \right )^2-k_{\textrm{in}}^2}}   e^{i k_{\textrm{in}}(x-x_{\textrm{in}})} 
\end{equation}
Using the parabolic approximation, one can simplify the phase terms as follows:
\begin{equation}
\label{psf4}
H(\mathbf{r},\mathbf{r}_{\textrm{in}},f) =e^{i2\pi f \left ( \frac{z}{c} -\frac{z_{\textrm{in}}}{c_0} \right) } \underbrace{ \frac{1}{2}  \int  d k_{\textrm{in}} \frac{e^{i \frac{k_{\textrm{in}}^2}{2\pi f/c} (z-c_0 z_{\textrm{in}}/c ) } }{\sqrt{\left (\frac{2\pi f}{c_0} \right )^2-k_{\textrm{in}}^2}}   e^{i k_{\textrm{in}}(x-x_{\textrm{in}})} }_{H^{(0)}_{\textrm{in}} \left (x-x_{\textrm{in}},\frac{c_0}{c} \left (z-z^{\star}_{\textrm{in}} \right ),f\right) }
\end{equation}

\section{Theoretical expression for the point spread function in presence of a reverberating plate}
\label{B}

In presence of the reverberating plate, each component of the focused reflection matrix (Eq.~\ref{decompose}) can be expressed as follows:
\begin{equation}
\label{Rbright}
\overline{\mathbf{R}}^{(l,m)}_{\Delta \mathbf{r}}(f) = \gamma_s \overline{B}(f) H^{(m)}_{\textrm{out}}(x_s,z_s,x_{\textrm{out}},z_{\textrm{out}},f)  H^{(l)}_{\textrm{in}}(x_{\textrm{in}},z_s,x_{\textrm{in}},z_{\textrm{in}},f). 
\end{equation}
where $H^{(l)}_{\textrm{in}}(x_s,z_s,x_{\textrm{in}},z_{\textrm{in}},f) $ and $H^{(m)}_{\textrm{out}}(x_s,z_s,x_{\textrm{out}},z_{\textrm{out}},f) $ stand for the $l^{\textrm{th}}$ and $m^{\textrm{th}}$ reverberation  order of the input and output point spread functions, respectively. The $l^{\textrm{th}}$ and $m^{\textrm{th}}$ reverberation order is associated with a number $2l$ and $2m$ of internal reflections inside the reverberating layer. 

Following the same rationale as in Appendix~\ref{A}, each reverberation order of the input point spread function can be expressed as follows:
\begin{equation}
\label{psfr1}
H^{(l)}_{\textrm{in}}(\mathbf{r},\mathbf{r}_{\textrm{in}},f) = \frac{1}{2} \int  d k_{\textrm{in}} \frac{e^{i (z-L)  \sqrt{ \left (\frac{2\pi f}{c} \right )^2-k_{\textrm{in}}^2} }e^{i (2l+1)L  \sqrt{ \left (\frac{2\pi f}{c_R} \right )^2-k_{\textrm{in}}^2} }e^{-i z_{\textrm{in}}  \sqrt{ \left (\frac{2\pi f}{c_0} \right )^2-k_{\textrm{in}}^2} }}{\sqrt{\left (\frac{2\pi f}{c_0} \right )^2-k_{\textrm{in}}^2}}   e^{i k_{\textrm{in}}(x-x_{\textrm{in}})} 
\end{equation}
Assuming a wave velocity model now equal to the phantom speed-of-sound ($c_0=c$), the last expression becomes
\begin{equation}
\label{psfr2}
H^{(l)}_{\textrm{in}}(\mathbf{r},\mathbf{r}_{\textrm{in}},f) = \frac{1}{2} \int  d k_{\textrm{in}} \frac{e^{i (z-L-z_{\textrm{in}})  \sqrt{ \left (\frac{2\pi f}{c} \right )^2-k_{\textrm{in}}^2} }e^{i (2l+1)L  \sqrt{ \left (\frac{2\pi f}{c_R} \right )^2-k_{\textrm{in}}^2} } }{\sqrt{\left (\frac{2\pi f}{c} \right )^2-k_{\textrm{in}}^2}}   e^{i k_{\textrm{in}}(x-x_{\textrm{in}})} .
\end{equation}
Using the parabolic approximation, one can simplify the phase terms  as follows:
\begin{equation}
\label{psf5}
H^{(l)}_{\textrm{in}}(\mathbf{r},\mathbf{r}_{\textrm{in}},f) =e^{i2\pi f \left ( \frac{z-L-z_{\textrm{in}}}{c} +\frac{(2l+1)L}{c_R} \right) } \underbrace{ \frac{1}{2}  \int  d k_{\textrm{in}} \frac{e^{i \frac{k_{\textrm{in}}^2}{2\pi f/c} (z-L- z_{\textrm{in}} +c_R (2l+1)L/c ) } }{\sqrt{\left (\frac{2\pi f}{c} \right )^2-k_{\textrm{in}}^2}}   e^{i k_{\textrm{in}}(x-x_{\textrm{in}})} }_{H^{(0)}_{\textrm{in}} \left(x-x_{\textrm{in}},z-z^{\star}_{\textrm{in}},f\right) }
\end{equation}
with
\begin{equation}
    z^{\star(l)}_{\textrm{in}}=z_{\textrm{in}}+L-(2l+1)L\frac{c_R}{c},
\end{equation}
the depth of the focusing plane for the $l^{\textrm{th}}$ reverberation order. As in Appendix \ref{A}, each reverberation component of the point spread function is thus the product of an axial propagation term which oscillates rapidly with frequency and a transverse focusing term that evolves slowly with frequency. In first approximation, the latter term can thus be replaced by its value at the central frequency $f_c$:
\begin{equation}
\label{psf5bis}
H^{(l)}_{\textrm{in}}(\mathbf{r},\mathbf{r}_{\textrm{in}},f) =\underbrace{e^{i2\pi f \left ( \frac{z-L-z_{\textrm{in}}}{c} +\frac{(2l+1)L}{c_R} \right) }}_{\textrm{axial propagation}}\underbrace{H^{(0)}_{\textrm{in}} \left(x-x_{\textrm{in}},z-z^{\star (l)}_{\textrm{in}},f_c\right) }_{\textrm{transverse focusing}}
\end{equation}
Note that an equivalent expression can be obtained for the reverberation components of the output PSF by replacing the index $l$ by index $m$:
\begin{equation}
\label{psf6}
H^{(m)}_{\textrm{out}}(\mathbf{r},\mathbf{r}_{\textrm{out}},f) =\underbrace{e^{i2\pi f \left ( \frac{z-L-z_{\textrm{out}}}{c} +\frac{(2m+1)L}{c_R} \right) }}_{\textrm{axial propagation}}\underbrace{H^{(0)}_{\textrm{out}} \left(x-x_{\textrm{out}},z-z^{\star (m)}_{\textrm{out}},f_c\right) }_{\textrm{transverse focusing}}
\end{equation}
with
\begin{equation}
    z^{\star(m)}_{\textrm{out}}=z_{\textrm{out}}+L-(2m+1)L\frac{c_R}{c}.
\end{equation}

Injecting Eqs.~\ref{psf5bis} and \ref{psf6} into Eq.~\ref{Rbright} leads to the following equation for each reflection matrix component, $\overline{{R}}^{(l,m)}(x_s,\Delta z,z_{\textrm{in}},f)$:
\begin{eqnarray}
\overline{{R}}^{(l,m)}(\Delta x,\Delta z,x_{\textrm{in}},z_{\textrm{in}},f) = &\gamma_s& \overline{B}(f) e^{i2\pi f \left ( \frac{2(z-L-z_{\textrm{in}})-\Delta z}{c} +\frac{(2l+2m+2)L}{c_R} \right) } \nonumber \\
&\times &H^{(0)}_{\textrm{out}}(\Delta x,z_s-z_{\textrm{in}}^{\star(m)} -\Delta z,f)  H^{(0)}_{\textrm{in}}(\Delta x,z_s-z_{\textrm{in}}^{\star (l)},f). 
\label{Rbright2}
\end{eqnarray}
The position of the scatterer $z_s$ is controlled by the time-of-flight expected for a scatterer at the targeted depth $z_{\textrm{in}}$ under a wave velocity model $c$: 
\begin{equation}
    z_s=z_{\textrm{in}}+L \left (1- \frac{c}{c_R} \right )
\end{equation}
Injecting this expression of $z_s$ into Eq.~\ref{Rbright2} and using the expressions of the focusing depths leads to the following expression of the focused $\mathbf{R}-$matrix coefficients in de-scanned coordinates:
\begin{equation}
   {R}^{(l,m)}(\Delta x,\Delta z,x_{\textrm{in}},z_{\textrm{in}},\tau) \propto \underbrace{H_{\textrm{in}}^{(0)}\left(0,\Delta z_l^{\star} ,f_c\right)}_{\textrm{constant}} \underbrace{H_{\textrm{out}}^{(0)}\left(\Delta x,\Delta z^{\star}_m-\Delta z ,f_c\right)}_{\textrm{transverse focusing}} \underbrace{\overline{B}(f) e^{-i2\pi f \left (\frac{\Delta z}{c} -\frac{2(l+m)L}{c_R} \right) }}_{\textrm{axial propagation}}.
   \label{Rbright3bis}
\end{equation}
with 
\begin{equation}
  \Delta z^{\star}_n = 2nL \frac{c_R}{c} + L \left ( \frac{c_R}{c} - \frac{c}{c_R} \right ),
\end{equation}
the focusing depth of the $n^{\textrm{th}}$ reverberation order in the de-scanned frame. An inverse Fourier transform of Eq.~\ref{Rbright3bis} leads to final expression of the focusing reflection matrix coefficients in the time domain for an isolated bright scatterer at targeted position $(x_{\textrm{in}},z_{\textrm{in}})$ :
\begin{equation}
   {R}^{(l,m)}(\Delta x,\Delta z,x_{\textrm{in}},z_{\textrm{in}},\tau) \propto \underbrace{H_{\textrm{in}}^{(0)}\left(0,\Delta z_l^{\star} ,f_c\right)}_{\textrm{constant}} \underbrace{H_{\textrm{out}}^{(0)}\left(\Delta x,\Delta z^{\star}_m-\Delta z ,f_c\right)}_{\textrm{spatial focusing}} \underbrace{B\left( \tau+ \frac{\Delta z} {c} - \frac{2(l+m)L}{c_R}\right)}_{\textrm{time focusing}}.
   \label{Rbright4}
\end{equation}
The ballistic component $(l=m=0)$ is expected to focus at a de-scan focusing depth $\Delta z^{\star}_0= L \left ( \frac{c_r}{c} - \frac{c}{c_r} \right )$ and focusing time $\tau^{\star}_0=-\Delta z^{\star}_0/c$.




\bibliography{references2}

\end{document}